\documentclass[aps,twocolumn,superscriptaddress,showpacs,pra,twoside,amssymb,amsmath]{revtex4}
\usepackage{txfonts}
\usepackage{type1cm}
\usepackage{graphicx}

\DeclareMathOperator*{\argmax}{argmax}

\begin{document}
\title{Security analysis of decoy state quantum key distribution incorporating finite statistics}

\author{Jun Hasegawa}
\affiliation{
Quantum Computation and Information Project, ERATO-SORST, Japan Science and Technology Agency, 
Daini Hongo White Building 201, 5-28-3 Hongo, Bunkyo-ku, Tokyo 113-0033, Japan
}
\affiliation{
Department of Computer Science, Graduate School of Information Science and Technology, 
the University of Tokyo, 
7-3-1 Hongo, Bunkyo-ku, Tokyo 113-0033, Japan
}
\author{Masahito Hayashi}
%\thanks{Corresponding author}
\affiliation{
Quantum Computation and Information Project, ERATO-SORST, Japan Science and Technology Agency, 
Daini Hongo White Building 201, 5-28-3 Hongo, Bunkyo-ku, Tokyo 113-0033, Japan
}
\author{Tohya Hiroshima}
\affiliation{
Quantum Computation and Information Project, ERATO-SORST, Japan Science and Technology Agency, 
Daini Hongo White Building 201, 5-28-3 Hongo, Bunkyo-ku, Tokyo 113-0033, Japan
}
\affiliation{
Nanoelectronics Research Laboratories, NEC Corporation,
34 Miyukigaoka, Tsukuba 305-8501, Japan
}
\author{Akihisa Tomita}
\affiliation{
Quantum Computation and Information Project, ERATO-SORST, Japan Science and Technology Agency, 
Daini Hongo White Building 201, 5-28-3 Hongo, Bunkyo-ku, Tokyo 113-0033, Japan
}
\affiliation{
Nanoelectronics Research Laboratories, NEC Corporation,
34 Miyukigaoka, Tsukuba 305-8501, Japan
}

%%%%%%%%%%%%%%%%%%%%%%%%%%%%%%%%%%%%%%%%%%%%%%%%%%%%%%%%%%%%%%%%%%%%%%%%%%%%%%%
\begin{abstract}
 Decoy state method quantum key distribution (QKD) is one of the
 promising practical solutions to BB84 QKD with coherent light pulses.
 In the real world, however, statistical fluctuations with
 the finite code length cannot be negligible, and
 the securities of theoretical and experimental researches of the decoy
 method state QKD so far are based on the asymptotic GLLP's formula
 which guarantees only that the limit of eavesdropper's information becomes
 zero as the code length approaches infinity.
 In this paper, we propose a substantially improved decoy state QKD
 in the framework of the finite code length and
 derive the upper bound of eavesdropper's information in the finite code length
 decoy state QKD with arbitrary number of decoy states of different
 intensities incorporating the finite statistics.
 We also show the performance of our decoy QKD and optimal values of
 parameters by numerical simulation.
\end{abstract}
%03.67.-a 	Quantum information
%03.67.Dd 	Quantum cryptography
%03.67.Hk 	Quantum communication
%%%%%%%%%%%%%%%%%%%%%%%%%%%%%%%%%%%%%%%%%%%%%%%%%%%%%%%%%%%%%%%%%%%%%%%%%%%%%%%
\maketitle

\section{Introduction} %\label{Introduction}

%Quantum cryptography systems such as quantum key distribution QKD system have been attracting increasing attention worldwide as they guarantee unconditional security against future advances in eavesdropping technology. Since the first protocol proposed in 1984 [1], QKD now steps forward from the proof of principle to the validation of the practical feasibility. Nevertheless, the QKD technology should respond to the challenges from the real-world such as security proof under the practical setting. The security of the current QKD protocol is dependent on ideal conditions, including the use of a genuine single photon source and/or unlimited computational resources. Statistical fluctuation in the observed data will also affect the accuracy of parameter estimation to define eavesdropper's (Eve's) performances. Therefore, there has been increasing desire to develop a QKD protocol that could guarantee security even without these ideal conditions.

Quantum key distribution (QKD) was originally proposed by Bennett and Brassard in 1984 \cite{BB84} as a protocol, by which two parties, Alice and Bob, share secret keys by using a quantum communication channel as well as a public classical channel \cite{QKD_Review}.
A remarkable feature is its unconditional security \cite{Mayers,Security,ILM}; it is guaranteed by the fundamental laws of quantum mechanics and thereby QKD provides the unconditionally secure communication system.
This is a triumph of quantum mechanics and quantum information science \cite{NC,Hayashi_Book} over the conventional cryptographic systems.
In the practical setting of optical communication, however, it is the almost only option to substitute qubits in the original BB84 QKD protocol with heavily attenuated laser pulses because the perfect single photon emitting devices are not available in the current technology.
Such laser pulses - the phase randomized weak coherent states - contains inevitably the multi-photon states at small but finite probability, which gives a malicious eavesdropper (Eve) a chance to obtain some amount of information on the shared keys by a photon-number-splitting attack \cite{PNS}.
Gottesman-Lo-L\"{u}tkenhaus-Preskill (GLLP) showed, however, that it is still possible to obtain unconditionally secret key by BB84 protocol with such imperfect light sources, although the key generation rate and distances are very limited \cite{GLLP}.

The recently proposed decoy state method \cite{Hwang,Wang,LMC,Ma} is one of the promising practical solutions to BB84 QKD with coherent light pulses, in which several coherent state pulses with different intensities are used.
Such optical pulses with different intensities have different photon number statistics.
This simple fact equips Alice and Bob with a countermeasure against Eve.
The original idea of the decoy state QKD is due to Hwang \cite{Hwang}.
So far, several experimental demonstrations of decoy state QKD have been reported \cite{ZQMLQ,Fiber,Free-Space,PZYGMYZYWP,YSS}.
In most cases, the security analysis is based on the GLLP's asymptotic arguments, whereas, in the practical setting, the code length is finite so that the asymptotic argument is no longer valid and the {\it unconditional security} is actually not guaranteed any more.
The security analysis of QKD with the finite code size must incorporate the statistical fluctuations of the observed quantities \cite{Hayashi_PRA}.
Although several authors \cite{Wang,Ma,Fiber,HEHN} have considered the influence of statistical fluctuations on the decoy state QKD with finite code length, 
what all of them have done is limited to the re-adjustment of parameters
of the asymptotic GLLP's formula. %for the secure key generation rate.
Such an {\em ad hoc} treatment cannot be justified to claim the
unconditional security
%because such a rate guarantees only that the limit of eavesdropper's
%information
%becomes zero as the code length approaches infinity.
because the asymptotic GLLP's formula for the secure key generation rate
provides us little knowledge about
the eavesdropper's information when the finite code length is given.
Therefore incorporating statistical fluctuations
to the parameters of the asymptotic GLLP's formula cannot
guarantee the security of the QKD protocols with the finite code length
even if values of these parameters are exactly known.
Suppose that the asymptotic rate of sacrifice bits
needed for the secure final private key is $R$.
When the code length is $n$,
assigning $nR$ to the number of sacrifice bits cannot
ensure how secure the final key is at all.
Because the asymptotic argument can only guarantee that
the limit of eavesdropper's information becomes zero
when the rate of sacrifice bits is greater than $R$,
and without the speed of the convergence
the increasing amount of sacrifice bits from $nR$,
which is needed for the secure key with the code length $n$,
cannot be estimated.
Thus we must consider statistical fluctuations
to eavesdropper's information formula with the finite code length $n$.
Several upper bounds of eavesdropper's information with $n$
have been provided~\cite{Mayers,ILM,Hayashi_Tight},
and especially
Hayashi's formula is simple and gives better key generation rate
than the others, some parameters of which
cannot be directly obtained from observed quantities
and are needed for estimation incorporating statistical fluctuations.

In this paper, we propose a substantially improved decoy state QKD in
the framework of finite code length~\cite{Hayashi_Tight}
by using the convex expansion formulas of
weak coherent states~\cite{Wang,Hayashi_Tight}
and
derive an eavesdropper's information formula in the finite code length
decoy state QKD with arbitrary number of decoy states of different
intensities incorporating the finite statistics.
We also show the performance of our decoy QKD and optimal values of
parameters by numerical simulation.

The rest of this paper is organized as follows:
In Section~\ref{Protocol}, we begin by describing our decoy method QKD
protocol.
In Section~\ref{General}, we show Eve's information considering
dark counts.
We next explain how to estimate Eve's information
 in Section~\ref{Eve} and random variables for describing the system
 in Section~\ref{Random}, 
 followed by
the estimation incorporating statistical fluctuations in
Section~\ref{Reverse}.
We then demonstrate our numerical results
of decoy method QKD protocol in Section~\ref{Numerical}
and finally summarize our results and discuss future work
in section~\ref{Conclusions}.

\section{Protocol} \label{Protocol}
First of all, we describe our protocol \cite{Hayashi_Tight}.
We fix the size $N$ of our code,
the number $N^{\prime }$ of sent pulses, 
the maximum number $\overline{N}$ and the minimum number $\underline{N}$
of the size of a final key.
We use $k+1$ different intensities or mean photon numbers $\mu _{0}=0<\mu _{1}<\ldots <\mu _{k}$ including vacuum ($\mu _{0} $) for the optical pulses.
Two conjugate bases ($+$ and $\times $) are treated separately so that $2k+1$ different pulses are involved in total.
The vacuum state ($i=0$) is sent at the probability $\overline{p}_{0}$ and the $\mu _{i} $ pulse with $\times $ ($+$) basis is sent at the probability $\overline{p}_{i}$ ($\overline{p}_{i+k}$) ($i=1,\ldots ,k$).
The pulse with the intensity $\mu _{i_{0}}$ ($\mu _{i_{0}+k}$) (the signal pulse) is used to distill a final secret key and the remainings (decoy pulses) are used just for estimation of Eve's attacks and/or the noise characteristics of quantum channel.

Before running the protocol, 
the probability $p_{D}$ of dark counts in the detector 
and the other (basis-dependent) system error probability $p_{S}$ ($\tilde{p}_{S}$) of the $\times$ ($+$) basis are measured in advance.
The probability $p_{S}$ or $\tilde{p}_{S}$ is the probability of errors
other than the transmission errors, that is, the error probability for
the noiseless channel.
We assume that the detector is a threshold detector and
the efficiency of the detector is independent of measurement bases~\cite{Hayashi_Tight}.

The protocol is as follows.
Alice randomly sends Bob a sequence of optical pulses of $k+1$ different intensities with randomly chosen basis.
After that, Bob performs a measurement in one of the two bases and they compare bases and keep the pulses with the common basis by communicating via public channel.
The number of sending pulses, received pulses, and pulses of the common basis are denoted by, respectively, $A_{i}$, $C_{i}$, and $E_{i}$ ($i=0,\ldots, 2k$).
Note that 
$\sum_{i=0}^{2k+1}A_{i}= N^{\prime }$.
The $E_{i} $ bit string of $i$th kind of pulse contains error bits, which will be detected by checking a portion of the bits (check bits).
To prepare check bits, they firstly perform the random permutation on $E_{i_{0}} $ and $E_{i_{0}+k} $ bit strings by sharing common random numbers via public channel.
Then, for $i=i_{0}$ and $i=i_{0}+k$, the first $N$ bit string is used as the raw key and the remaining $E_{i_{0}}-N$ and $E_{i_{0}+k}-N$ bit string are used as the check bits, while the whole $E_{i}$ bits are used as check bits for $i\neq i_{0},i_{0}+k$.
(If $E_{i_{0}}\leq N$ or $E_{i_{0}+k}\leq N$, then the protocol is aborted.)
The number of detected errors of $i$th kind of pulse is denoted by $H_{i}$ $(i=1,\ldots,2k)$.
From these quantities, they can evaluate the size of the final key guaranteeing the unconditional security.
If the evaluated final key size is not positive, the protocol is aborted again.
The size of final secret key of $+$ basis is computed as
\begin{equation}
N_{final}:=N\eta \left( \frac{H_{i_{0}+k}}{E_{i_{0}+k}-N}\right) -m(\mathcal{D%
}_{i},\mathcal{D}_{e}),
\end{equation}
and that of $\times$ basis is
\begin{equation}
\hat{N}_{final}:=N\eta \left( \frac{H_{i_{0}}}{E_{i_{0}}-N}\right) -\tilde{m}(\tilde{\mathcal{%
D}}_{i},\mathcal{D}_{e}),
\end{equation}
where $\eta (\cdot )$ denotes the error correcting coding rate and 
$m(\mathcal{D}_{i},\mathcal{D}_{e})$ and $\tilde{m}(\tilde{\mathcal{D}}_{i},\mathcal{D}_{e})$ 
represents the size of privacy amplification.
Here we abbreviate the initial data 
$(\mathbf{A},\pmb{\mu},p_{S},p_{D})$, 
$(\mathbf{A},\pmb{\mu},\tilde{p}_{S},p_{D})$
and the observed data 
$(\mathbf{C},\mathbf{E},\mathbf{H})$ to 
$\mathcal{D}_{i}$, $\tilde{\mathcal{D}}_{i}$, and $\mathcal{D}_{e}$, respectively, and
$\mathbf{A}=(A_{1},\ldots ,A_{2k})$, etc.
If $N_{final} < \underline{N}$ or $\hat{N}_{final} <\underline{N}$, 
they abort the protocol and go back to the first step.
Furthermore, if
$\overline{N}< N_{final}$ ($\overline{N}< \hat{N}_{final}$ ), 
they replace $m(\mathcal{D}_{i},\mathcal{D}_{e})$ 
[$\tilde{m}(\tilde{\mathcal{D}}_{i},\mathcal{D}_{e})$]
by $N\eta(H_{i_{0}+k}/(E_{i_{0}+k}-N))-\overline{N}$ 
[$N\eta(H_{i_{0}}/(E_{i_{0}}-N))-\overline{N}$].
Finally, they are left with $N$ bits error correction followed by privacy amplification 
to share the $N_{final}$ ($\hat{N}_{final}$) bit secret key of $+$ ($\times$) basis.

The error correction is performed as follows.
Suppose that Alice and Bob have, respectively, the random number sequences 
$X$ and $X^{\prime }$ of $N$ bits, which contain some errors.
The task is to distill the common random number sequence of $l+m$ bits with negligible errors.
In the forward error correction, 
they share $N \times (l+m)$ binary matrix $M_{e}$.
Alice generates other $l+m$ bits random number $Z$, 
and sends a bit sequence $M_{e} Z+X$ to Bob.
Then, Bob applies the decoding of the code $M_{e}$ to the bit sequence 
$M_{e} Z+X-X^{\prime }$ to extract $Z$.
On the other hand, in the reverse error correction,
Bob generates the random number sequence $Z$ of $l+m$ bits,
and sends $M_{e} Z+X^{\prime }$ to Alice.
Then, Alice applies the decoding code $M_{e}$ to the bit sequence 
$M_{e} Z+X^{\prime }-X$ to extract $Z$.
The error correction here corresponds to a part of the twirling operation so that 
their channel can be regarded as a Pauli channel 
from Alice (Bob) to Bob (Alice) in the forward (reverse) error correction \cite{Hayashi_Tight}.

In the privacy amplification, 
Alice and Bob share the final secret key of $l$ bits from $Z$ of $l+m$ bits.
More precisely, they first generate the same $l\times (l+m)$ binary matrix $M_p$ 
with
\begin{equation}
\mathrm{Prob} \{Z \in \mathrm{Im} M_p^T\}\le 2^{-m}
\end{equation}
for any non-zero $l+m$ bit sequence $Z$.
Subsequently, they generate the bit sequence $M_p Z$ of $l$ bits from $Z$ of $l+m$ bits.

Combining the error correction and the privacy amplification described above, 
the sequel of it is that Alice sends information by the code 
$\mathrm{Im} M_{e}/M_{e} (\mathrm{Ker} M_p)$.

\section{General upper bound for Eve's information on final key} \label{General}

In this section, we give an upper bound for the leakage information on the final key, 
which lays the foundation of the security analysis in Sec.~\ref{Reverse} \cite{Hayashi_Tight}.
Here, we confine ourselves to the discussion on the final key with $+$ basis.
Eve's attack can be described by the conditional distribution $\mathcal{P}$ of the Pauli action on the input state.
Hence, the average of Eve's information with respect to the final key is closely related to the error probability 
$P^{\mathcal{P}}_{ph,min,x|M_p,\mathcal{D}_{e},\mathrm{POS}}$ with the minimum distance decoding when information is sent with $\times$ basis and the code $\mathrm{Im} M_{e}/M_{e} (\mathrm{Ker} M_p)$, 
where $\mathrm{POS}$ is a random variable for the arrangement of different intensities and the position of check bits, and $x=\to$ ($\leftarrow$) refers to the forward (reverse) error correction.
The average 
$I^{\mathcal{P}}_{E,av,x}=\mathbb{E}^{\mathcal{P}}_{M_p,\mathcal{D}_{e},\mathrm{POS}} 
I^{\mathcal{P}}_{E,x|M_p,\mathcal{D}_{e},\mathrm{POS}}$ 
of Eve's information $I^{\mathcal{P}}_{E,x|M_p,\mathcal{D}_{e},\mathrm{POS}}$ 
is evaluated in terms of 
$P^{\mathcal{P}}_{ph,av,x}
=\mathbb{E}^{\mathcal{P}}_{M_p,\mathcal{D}_{e},\mathrm{POS}} 
P^{\mathcal{P}}_{ph,min,x|M_p,\mathcal{D}_{e},\mathrm{POS}}$  as
\begin{equation}
I^{\mathcal{P}}_{E,av,x} \le 
P^{\mathcal{P}}_{ph,av,x} (1+\overline{N}-\log P^{\mathcal{P}}_{ph,av,x}).
\label{Information_Inequality_1}
\end{equation}
Since the stochastic behavior of the random variables $\mathcal{D}_{e}$ depends on the conditional distribution $\mathcal{P}$, 
we denote the operation of taking the expectation with respect to 
$M_p,\mathcal{D}_{e},\mathrm{POS}$ 
by $\mathbb{E}^{\mathcal{P}}_{M_p,\mathcal{D}_{e},\mathrm{POS}}$.
Denoting Eve's state with respect to the final key $[Z]=M_p Z$ by $\rho^{E,x}_{[Z]}$, 
and its average state by $\overline{\rho}^{E,x}$, 
we obtain the following inequalities.
\begin{align}
\mathbb{E}^{\mathcal{P}}_{M_p,\mathcal{D}_{e},\mathrm{POS}} 
\min_{[Z]\neq[Z^{\prime }]}
F(\rho^{E,x}_{[Z]},\rho^{E,x}_{[Z^{\prime }]})
& \ge 1-2 P^{\mathcal{P}}_{ph,av,x}, \\
\mathbb{E}^{\mathcal{P}}_{M_p,\mathcal{D}_{e},\mathrm{POS}} 
\max_{[Z]\neq[Z^{\prime }]}
\|\rho^{E,x}_{[Z]}-\rho^{E,x}_{[Z^{\prime }]}\|_{1} &\le 4 P^{\mathcal{P}}_{ph,av,x}, \\
\mathbb{E}^{\mathcal{P}}_{M_p,\mathcal{D}_{e},\mathrm{POS}} 
\min_{[Z]}
F(\rho^{E,x}_{[Z]},\overline{\rho}^{E,x})
& \ge 1-2 P^{\mathcal{P}}_{ph,av,x}, \\
\intertext{and}
\mathbb{E}^{\mathcal{P}}_{M_p,\mathcal{D}_{e},\mathrm{POS}} 
\max_{[Z]}
\|\rho^{E,x}_{[Z]}-\overline{\rho}^{E,x}\|_{1} &\le 4 P^{\mathcal{P}}_{ph,av,x}.
\end{align}
Here, we have omitted the dependence of $\rho^{E,x}_{[Z]}$ on $M_p$, $\mathcal{D}_{e}$, and $\mathrm{POS}$.
Next, let $P^{\mathcal{P}}_{succ,x|M_p}$ be the probability that Eve acquires perfectly information on the final key when she performs the optimal measurement after the privacy amplification.
Then, 
\begin{eqnarray}
&\mathbb{E}^{\mathcal{P}}_{M_p,\mathcal{D}_{e},\mathrm{POS}} 
P^{\mathcal{P}}_{succ,x|M_p} \nonumber \\
\le &
\Bigl(
\sqrt{P^{\mathcal{P}}_{ph,av,x}}
\sqrt{\strut{1-2^{-\underline{N}}}}+
\sqrt{1-P^{\mathcal{P}}_{ph,av,x}}
\sqrt{\strut{2^{-\underline{N}}}}
\Bigr)^{2}
\end{eqnarray}
Here, we have again omitted the dependence of $P^{\mathcal{P}}_{succ,x|M_p}$ on $M_p$, $\mathcal{D}_{e}$, and $\mathrm{POS}$.
Now, it is evident that the evaluation of $P^{\mathcal{P}}_{ph,av,x}$
plays an essential role in the security analysis.

We start by grouping detected pulses into six parts according to which state (vacuum, single photon, or multi-photon) is actually sent by Alice and whether or not the detection is normal, i.e., it is {\em not} due to the dark counts.
We define $J^{i}$ ($J^{3+i}$) as the number of pulses detected normally (by dark counts) under the condition that the state sent is vacuum ($i=0$), single photon ($i=1$), or multi-photon ($i=2$) states.
For example, $J^{3}$ represents the number of pulses detected by dark
counts when the state sent by Alice is the vacuum.
We regard the simultaneous event of a dark count and a normal count
as a dark count.
This is because the collision of both photons causes the information
loss of the normal count.
Let $t$ be the number of pulses or bits with transmission (phase) error in $\times $ basis among $J^{1}$ bits.
This is also a random variable.
Then, by denoting the expectation with respect to the random variables $t$ and 
$\mathbf{J}=(J^{0},\ldots,J^{5})$ by $\mathbb{E}_{t,\mathbf{J}}^{\mathcal{P}}$, 
$P^{\mathcal{P}}_{ph,av,x}$ can be evaluated as
\begin{align}
P_{ph,av,\rightarrow }^{\mathcal{P}}\le & \mathbb{E}_{t,\mathbf{J},\mathcal{D}_{e},%
\mathrm{POS}}^{\mathcal{P}}2^{-\left[ m(\mathcal{D}_{i},\mathcal{D}%
_{e})-J^{1}\overline{h}(t/J^{1})-J^{2}-J^{4}-J^{5}\right] _{+}}  \nonumber \\
=& \mathbb{E}_{t,\mathbf{J},\mathcal{D}_{e},\mathrm{POS}}^{\mathcal{P}%
}2^{-\left[ m(\mathcal{D}_{i},\mathcal{D}_{e})-N+J^{1}(1-\overline{h}%
(t/J^{1}))+J^{0}+J^{3}\right] _{+}},
\label{Phase_Error_1} \\
\intertext{and}
P_{ph,av,\leftarrow }^{\mathcal{P}}\le & \mathbb{E}_{t,\mathbf{J},\mathcal{D}%
_{e},\mathrm{POS}}^{\mathcal{P}}2^{-\left[ m(\mathcal{D}_{i},\mathcal{D}%
_{e})-J^{1}\overline{h}(t/J^{1})-J^{0}-J^{2}\right] _{+}}  \nonumber \\
=& \mathbb{E}_{t,\mathbf{J},\mathcal{D}_{e},\mathrm{POS}}^{\mathcal{P}%
}2^{-\left[ m(\mathcal{D}_{i},\mathcal{D}_{e})-N+J^{1}(1-\overline{h}%
(t/J^{1}))+J^{3}+J^{4}+J^{5}\right] _{+}},
\label{Phase_Error_2}
\end{align}
where $[z]_{+}=\max \{0,z\}$ and $\overline{h}(x)$ is defined by
\begin{equation}
\overline{h}(x):=
\left\{
\begin{array}{cl}
-x \log_{2} x -(1-x)\log_{2} (1-x) & \hbox{ if } x \in [0,1/2] \\
1 & \hbox{ if }x \in ( 1/2, 1].
\end{array}
\right.
\end{equation}
In the actual system, the random variables $t$ and $\mathrm{J} $ cannot be identified exactly.
They are estimated from the measured values $\mathcal{D}_{e}$, by which the size of sacrifice bits $m(\mathcal{D}_{i},\mathcal{D}_{e})$ is determined.
It is of crucial importance to determine $m(\mathcal{D}_{i},\mathcal{D}_{e})$ such that the average error probability $P^{\mathcal{P}}_{ph,av,x}$ is less than a given value for any attack $\mathcal{P}$.
The statistical fluctuation of $\mathcal{D}_{e}$ is properly taken into account in the computation of $m(\mathcal{D}_{i},\mathcal{D}_{e})$.
As for the attack $\mathcal{P}$, it is sufficient to treat the extremal points, in which these random variables can be described by the combination of multi-hypergeometric distributions.
All random variables concern our problem are listed in Sec.~\ref{Random}.

\section{Eve's strategy and its estimation} \label{Eve}

Suppose that Eve can distinguish the different number states.
A na\"ive way to describe Eve's attacks is to associate each number state with the corresponding parameters describing Eve's attacks.
This is, however, a formidable task because the infinite number of unknown parameters are involved.
In order to avoid such a difficulty, one of the authors \cite{Hayashi_Asymptotic} introduced a convex expansion of the phase-randomized coherent state 
$\sum_{n=0}^{\infty }e^{-\mu }\mu ^{n}|n\rangle \langle n|/n!$
in terms of vacuum, single-photon, and multi-photon states.
Here, we define the multi-photon states $\sigma _{l}$ ($l=2,\ldots, k+1$) as
\begin{equation}
\sigma _{l}:=
\frac{1}{\Omega_{l}}
\sum_{n=l}^\infty
\frac{\gamma_{l,n}}{n!}
|n \rangle \langle n|,
\label{multi-photon}
\end{equation}
where
\begin{equation}
\Omega_{l}:=\sum_{n=l}^\infty\frac{\gamma_{l,n}}{n!},
\end{equation}
and
\begin{equation}
\gamma _{l,n}:=\sum_{j=1}^{l-1}\frac{\mu_{j}^{n-2}}{\prod_{t=1,t\neq j}^{l-1}(\mu_j-\mu_t)},
%\gamma _{l,n}:=\sum_{j=1}^{l-1}\frac{(\mu _{l-1}-\mu _{l-2})\cdots (\mu
%_{l-1}-\mu _{1})\mu _{l-1}^{2}\mu _{j}^{n-2}}{(\mu _{j}-\mu _{l-1})\cdots
%(\mu _{j}-\mu _{j+1})(\mu _{j}-\mu _{j-1})\cdots (\mu _{j}-\mu _{1})},
\end{equation}
with $\mu_{1}< \mu_{2}< \cdots < \mu_k$.
Note that $\sigma _{l} $ [Eq.~(\ref{multi-photon})] are {\em bona fide} states, i.e., 
$\sigma _{l}\geq 0$ and $\mathrm{Tr}\sigma _{l}=1$.
It is easy to see that the state 
$\sum_{n=0}^{\infty }e^{-\mu }\mu ^{n}|n\rangle \langle n|/n!$ 
can be expressed as a convex combination of 
$\left| 0\right\rangle \left\langle 0\right| $,
$\left| 1\right\rangle \left\langle 1\right| $, and $\sigma _{l} $;
\begin{eqnarray}
&&e^{-\mu _{i}}\sum_{n=0}^{\infty }\frac{\mu _{i}^{n}}{n!} | n \rangle
 \langle n |  \nonumber \\
 &=&e^{-\mu_i} \left( |0\rangle \langle 0|
	       + \mu _{i}|1\rangle \langle 1|
	       + \sum_{n=2}^{i+1}\mu_i^2 \prod_{t=1}^{n-2}(\mu_i-\mu_t)\Omega _{n}\sigma _{n} \right).
%&=&e^{-\mu _{i}}|0\rangle \langle 0|+\mu _{i}e^{-\mu _{i}}|1\rangle \langle
%1|  \nonumber \\
%&&+e^{-\mu _{i}}\sum_{n=2}^{i+1}\frac{\mu _{i}^{2}(\mu _{i}-\mu _{1})\cdots
%(\mu _{i}-\mu _{n-2})}{\mu _{n-1}^{2}(\mu _{n-1}-\mu _{1})\cdots (\mu
%_{n-1}-\mu _{n-2})}\Omega _{n}\sigma _{n}.
\end{eqnarray}
Here, coefficients
\begin{equation*}
 e^{-\mu_i}\mu_i^2 \prod_{t=1}^{n-2}(\mu_i-\mu_t)\Omega _{n}
%e^{-\mu _{i}}\frac{\mu_{i}^{2} (\mu_{i}-\mu_{1})\cdots(\mu_{i}-\mu_{n-2})}
%{\mu_{n-1}^{2} (\mu_{n-1}-\mu_{1})\cdots(\mu_{n-1}-\mu_{n-2})}
%\Omega_n
\end{equation*}
are positive.

Now, we adopt the worst case scenario on Eve's attacks.
Namely, we assume that Eve can distinguish vacuum state 
($\rho _{0}=\left| 0\right\rangle \left\langle 0\right| $), 
single photon state 
($\rho _{1}=\left| 1\right\rangle \left\langle 1\right| $), 
multi-photon states with $\times $ basis $\rho _{j}=\sigma _{j}$ ($j=2, \ldots, k+1$) and those with $+ $ basis $\rho _{k+j}=\sigma _{j}$ ($j=k+2, \ldots, 2k+1$).
The number of emitted $j$th state $\rho _{j} $ ($j=0, \ldots, 2k+1$) is denoted by $B^{j}$; 
$\mathbf{B}=(B^{j})$.
According to the values of $\mathbf{B}$, Eve can do the following attacks;
Eve tricks Bob into detecting $j$th state with ratio $q^{j}(\mathbf{B})$ \cite{comment1} and causes phase errors with ratio $r^{j}(\mathbf{B})$ for $j$th state ($j=1,2,\ldots,k+1$) and bit errors with ratio $\tilde{r}^{j}(\mathbf{B})$ for $j$th state ($j=1,k+2,\ldots,2k+1$).
The quantities $r^{j}(\mathbf{B})$ and $\tilde{r}^{j}(\mathbf{B})$ describe the transmission errors.
In the following, we focus on the final secret key of $\times $ basis and write $q^{j}$ and $r^{j}$ ($\tilde{r}^{j}$) instead of $q^{j}(\mathbf{B})$ and $r^{j}(\mathbf{B})$ [$\tilde{r}^{j}(\mathbf{B})$].
Note that $q^{j}$ is the rate of detection with the exclusion of dark counts.
Since the state $\rho_j$ with the $\times$ basis is, in general, different from that with the $+$ basis, 
the parameters $q^{j}$ do not necessarily coincide with $q^{j+k}$ ($j=1,\ldots,k$).
The generating probability of each state can be described by the matrix 
$(P_{k:i}^{j})_{i=0, \ldots, 2k, j= 0, \ldots, 2k+1}$ 
defined by
\begin{equation} \label{Pk}
P_{k}:=\begin{pmatrix}
1 &0& 0 & 0 \\
Y &Z& X & 0 \\
Y &Z& 0 & X
\end{pmatrix}
\end{equation}
where $Y$ and $Z$ are $k$-dimensional vectors such that
$Y_{i} =e^{-\mu_{i}} $ and $Z_{i} =\mu_{i} e^{-\mu_{i}} $ and 
the $k \times k$ matrix $X$ is given by
\begin{equation}
X_{i}^{j}:=
\left\{
\begin{array}{cl}
 \mu_i^2\prod_{t=1}^{j-1}(\mu_i-\mu_t)e^{-\mu_i}\Omega_{j+1}
%\frac{\mu_{i}^{2} (\mu_{i}-\mu_{1})\cdots(\mu_{i}-\mu_{j-1})}
%{\mu_{j}^{2} (\mu_{j}-\mu_{1})\cdots(\mu_{j}-\mu_{j-1})}
%e^{-\mu_{i}}\Omega_{j+1}
& j= 1,\ldots, i \\
0 & j = i+1,\ldots, k,
\end{array}
\right.
\end{equation}
for $i=1,\ldots, k$.
For example, the matrix $P_k$ for $k=3$ is given by
\begin{widetext}
\begin{eqnarray}
P_{3}&=&\begin{pmatrix}
1 & 0 & 0 & 0 & 
0 & 0 & 0 & 0 \\
e^{-\mu_{1}} & \mu_{1} e^{-\mu_{1}} & e^{-\mu_{1}}\mu_1^2\Omega_{2}  & 0 &
       0 & 0 & 0 & 0\\
e^{-\mu_{2}} & \mu_{2} e^{-\mu_{2}} & e^{-\mu_{2}}\mu_2^2\Omega_{2}
       & e^{-\mu_{2}}\mu_2^2(\mu_2-\mu_1)\Omega_{3}
       & 0 & 0 & 0 & 0\\
e^{-\mu_{3}} & \mu_{3} e^{-\mu_{3}} & e^{-\mu_{3}}\mu_3^2\Omega_{2}
       & e^{-\mu_{3}}\mu_3^2(\mu_3-\mu_1)\Omega_{3}
       & * & 0 & 0 & 0 \\
e^{-\mu_{1}} & \mu_{1} e^{-\mu_{1}} & 0 & 0 & 0 &
       e^{-\mu_1}\mu_1^2\Omega_2 & 0 & 0\\
e^{-\mu_{2}} & \mu_{2} e^{-\mu_{2}} & 0 & 0 & 0
	 & e^{-\mu_2}\mu_2^2\Omega_2
	 & e^{-\mu_{2}} \mu_2^2 (\mu_2-\mu_1) \Omega_{3} & 0 \\
e^{-\mu_{3}} & \mu_{3} e^{-\mu_{3}} & 0 & 0 & 0
	 & e^{-\mu_3}\mu_3^2\Omega_2
	 & e^{-\mu_{3}} \mu_3^2 (\mu_3-\mu_1) \Omega_{3}
       & *
%P_{3}=\begin{pmatrix}
%1 & 0 & 0 & 0 & 
%0 & 0 & 0 & 0 \\
%e^{-\mu_{1}} & \mu_{1} e^{-\mu_{1}} & \Omega_{2} e^{-\mu_{1}}& 0 &
%0 & 0 & 0 & 0\\
%e^{-\mu_{2}} & \mu_{2} e^{-\mu_{2}} & \Omega_{2}\frac{\mu_{2}^{2}}{\mu_{1}^{2}}e^{-\mu_{2}}
%& \Omega_{3} e^{-\mu_{2}}& 0 & 0 & 0 & 0\\
%e^{-\mu_{3}} & \mu_{3} e^{-\mu_{3}} & \Omega_{2}\frac{\mu_{3}^{2}}{\mu_{1}^{2}}e^{-\mu_{3}}
%& \Omega_{3} \frac{\mu_{3}^{2}(\mu_{3}-\mu_{1})}{\mu_{2}^{2}(\mu_{2}-\mu_{1})}e^{-\mu_{3}}
%& \Omega_{4} e^{-\mu_{3}}& 0 & 0 & 0 \\
%e^{-\mu_{1}} & \mu_{1} e^{-\mu_{1}} & 0 & 0 & 0 & \Omega_{2} e^{-\mu_{1}}& 0 & 0\\
%e^{-\mu_{2}} & \mu_{2} e^{-\mu_{2}} & 0 & 0 & 0 &
%\Omega_{2}\frac{\mu_{2}^{2}}{\mu_{1}^{2}}e^{-\mu_{2}}& \Omega_{3} e^{-\mu_{2}}& 0 \\
%e^{-\mu_{3}} & \mu_{3} e^{-\mu_{3}} & 0 & 0 & 0 &
%\Omega_{2}\frac{\mu_{3}^{2}}{\mu_{1}^{2}}e^{-\mu_{3}}& 
%\Omega_{3} \frac{\mu_{3}^{2}(\mu_{3}-\mu_{1})}{\mu_{2}^{2}(\mu_{2}-\mu_{1})}e^{-\mu_{3}}
%& \Omega_{4}e^{-\mu_{3}}
\end{pmatrix}, \\
 * &=& e^{-\mu_{3}}(\mu_3-\mu_1)(\mu_3-\mu_2) \Omega_{4}. \notag
\end{eqnarray}
\end{widetext}

Now, the expectation of the detection probability $p_{i}= C_{i}/A_{i} $ is expressed as
\begin{equation}
\mathbb{E} (p_{i}) = \Xi _{i}^{(1)}(\mathbf{q}) :=
\sum_{j=0}^{2k+1} P_{i}^{j} q^{j}+p_{D},
\label{p}
\end{equation}
for $i=0,\ldots,2k$ 
and the expectation of detected error probability is written as
\begin{equation}
\mathbb{E} (s_{i} p_{i}) = \Xi _{i}^{(2)}(\mathbf{q},\mathbf{r}) :=
P_{i}^{1} q^{1} {r^{1}}^{\prime }
+\sum_{j=2}^{k+1} P_{i}^{j} q^{j} r^{j}+\frac{1}{2} (P_{i}^{0} q^{0} + p_{D}),
\label{sp}
\end{equation}
for $i=1,2,\ldots,k$,
where 
$s_{i}= H_{i}/E_{i}$ ($i \neq i_{0}$) and
$s_{i_{0}}= H_{i_{0}}/(E_{i_{0}}-N)$.
In Eq.~(\ref{sp}), 
${r^{1}}^{\prime }= (1-p_{S})r^{1}+p_{S}(1-r^{1})$.
Since we cannot uniquely determine the parameters $q^{j}$ and $r^{j}$ from 
Eqs.~(\ref{p}) and (\ref{sp}) even when 
$\mathbb{E} (p_{i})=p_{i}$ and $\mathbb{E} (s_{i} p_{i})= s_{i} p_{i}$, 
we fix $\frac{q^{k+1}+q^{2k+1}}{\sqrt{2}}$ and $r^{k+1}$ 
as $x\in[0,\sqrt{2}(1-p_{D})]$ and $y\in [0,1]$ to obtain
\begin{equation}
\hat{q}^{j}_{x}=\left\{ 
\begin{array}{cl}
p_{0}-p_{D} & j=0 \\ 
\xi ^{j} & j=1,\ldots ,k,k+2,\ldots ,2k \\ 
\frac{1}{\sqrt{2}}\left( x+\xi ^{k+1}\right)  & j=k+1 \\ 
\frac{1}{\sqrt{2}}\left( x-\xi ^{k+1}\right)  & j=2k+1,
\end{array}
\right.
\label{q}
\end{equation}
and
\begin{equation}
\hat{r}_{x,y}^{j}=\left\{ 
\begin{array}{cl}
\frac{1}{1-2p_{S}}\left( \zeta ^{1}-p_{S}\right)  & j=1 \\ 
\zeta ^{j} & j=2,\ldots ,k \\ 
y & j=k+1,
\end{array}
\right.
\label{r}
\end{equation}
with
\begin{eqnarray}
\zeta ^{j} &:=&\frac{1}{\hat{q}_{x}^{j}}\sum_{i=1}^{k}(P_{k\times
k}^{-1})_{j}^{i} \nonumber \\
&&\times \left\{ s_{i}p_{i}-\frac{1}{2}[P_{i}^{0}(p_{0}-p_{D})+p_{D}%
]-P_{i}^{k+1}\hat{q}_{x}^{k+1}y\right\},
\label{zeta}
\end{eqnarray}
where $X_{l \times l}$ presents the $l \times l$ submatrix 
$(X_{i,j})_{1\le i,j\le l}$
for a given rectangular matrix $X=(X_{i,j})_{0\le i\le a,0 \le j \le b}$,
and $\xi ^{j}$ ($j=1,\ldots,2k$) in Eq.~(\ref{q}) is defined as
\begin{equation}
\xi ^{j}:=\sum_{i=1}^{2k}(\overline{P}_{2k\times
2k}^{-1})_{j}^{i}\left[ p_{i}-p_{D}-\overline{P}_{i}^{0}(p_{0}-p_{D})-x%
\overline{P}_{i}^{2k+1}\right],
\label{xi}
\end{equation}
where
\begin{equation}
\overline{P}_{i}^{j}:=
\sum_{j^{\prime }=0}^{2k+1 }P_{i}^{j^{\prime }} Q^{j^{\prime },j},
\end{equation}
with
\begin{eqnarray}
Q^{j^{\prime },j} := \left\{
\begin{array}{cl}
-\frac{1}{\sqrt{2}} & j^{\prime }= k+1, j= 2k+1 \\
\frac{1}{\sqrt{2}} & j^{\prime }= 2k+1, j= 2k+1 \\
\frac{1}{\sqrt{2}} & j^{\prime }= 2k+1 \hbox{ or } k+1, j= k+1 \\
\delta^{j^{\prime },j}& \hbox{ Otherwise.}
\end{array}
\right.
\end{eqnarray}

When the true values $q^{j}$ and $r^{j}$ are close to zero, say, 
the linear estimators given above often take on negative values due to statistical fluctuations.
The maximal likelihood estimation provides an alternate solution free from such a drawback, which is given by
\begin{widetext}
\begin{eqnarray}
\left( \mathbf{q}_{x,y,(B_{i}^{j})}^{ML},\mathbf{r}_{x,y,(B_{i}^{j})}^{ML}%
\right)  &=&\argmax _{\mathbf{q},\mathbf{r}:q^{k+1}+q^{2k+1}=\sqrt{2}%
x,r^{k+1}=y}\left\{ \sum_{i=0}^{2k}\left[ C_{i}\log \Xi _{i}^{(1)}(\mathbf{q}%
)+(A_{i}-C_{i})\log (1-\Xi _{i}^{(1)}(\mathbf{q}))\right] \right.   \nonumber
\\
&&\left. +\sum_{i=1}^{k}H_{i}\log \frac{\Xi _{i}^{(2)}(\mathbf{q},\mathbf{r})%
}{\Xi _{i}^{(1)}(\mathbf{q})}+\sum_{i=1}^{k}(E_{i}-\delta
_{i,i_{0}}N-H_{i})\log \left( 1-\frac{\Xi _{i}^{(2)}(\mathbf{q},\mathbf{r})}{%
\Xi _{i}^{(1)}(\mathbf{q})}\right) \right\} ,
\end{eqnarray}
\end{widetext}
with 
$0\leq q^{j}\leq 1-p_{D}$ and $0\leq r^{j}\leq 1$.
If the above linear estimators Eqs.~(\ref{q}) and (\ref{r}) are in the range 
$0\leq \hat{q}_{x}^{j}\leq 1-p_{D}$ and $0\leq \hat{r}_{x,y}^{j}\leq 1$, 
then they coincide with the corresponding maximum likelihood estimators.

\section{Random variables describing the system} \label{Random}

To compute the size of final secret key with the finite statistics due to the finite code length, several stochastic variables must be incorporated properly.
In this section, we describe random variables with their means and (co)variances, which are used in the computation of the sacrifice key size (Sec.~\ref{v}).

Firstly, we define $B^{j}_{i}$ as the number of the emitted state $\rho_j$ given that the $i$th kind of pulse is sent ($i=0,\ldots,2k$; $j=0,\ldots,2k+1$).
They obey the following multi-nomial distribution.
\begin{equation}
P(B^{0}_{i},\ldots, B^{2k+1}_{i})=
(P^{0}_{i})^{B^{0}_{i}}
\cdots (P^{2k+1}_{i})^{B^{2k+1}_{i}}
\frac{A_{i}!}{B^{0}_{i}! \cdots B^{2k+1}_{i}!}.
\end{equation}
Note that $A_{i} = \sum_{j=0}^{2k+1} B^{j}_{i}$ and $B^{j} = \sum_{i=0}^{2k} B^{j}_{i}$.
The mean of $B^{j}_{i}$ is $P^{j}_{i}A_{i}$.

Next, we define $C^{j}_{i}$ as the number of $i$the kind of pulse detected normally under the condition that the emitted states is $\rho_j$ ($i=0,\ldots,2k$; $j=0,\ldots,2k+1$).
Note that the dark counts are not included in the detection events.
The contribution of dark counts is expressed as $C^{-1}_{i}$ ($j=-1$).
These stochastic variables obey
\begin{equation}
P(C^{j}_{0},C^{j}_{1},\ldots,C^{j}_{2k})=
\frac{\binom{B^{j}_{0}}{C^{j}_{0}}
\binom{B^{j}_{1}}{C^{j}_{1}}
\cdots
\binom{B^{j}_{2k}}{C^{j}_{2k}}
}
{\binom{\sum_{i=0}^{2k} B^{j}_{i}}{q^{j} \sum_{i=0}^{2k} B^{j}_{i}}},
\end{equation}
for $j \ne -1$ 
and
\begin{equation}
P(C^{-1}_{i})=p_{D}^{C^{-1}_{i}}(1-p_{D})^{A_{i}-C^{-1}_{i}}
\binom{A_{i}}{C^{-1}_{i}}.
\end{equation}
Note that $\sum_{i=0}^{2k} C^{j}_{i}= q^{j} \sum_{i=0}^{2k} B^{j}_{i}$ and
$C_{i} = \sum_{j=-1}^{2k+1} C^{j}_{i}$.
The means $\mathbb{E} C^{j}_{i}$ are given by $q^{j} B^{j}_{i}$ 
for $j \ne -1$ and $p_{D} A_{i}$ for $j=-1$.
Deviations $\Delta ^{\prime } C^{j}_{i} =  C^{j}_{i} - \mathbb{E} C^{j}_{i}$ satisfy
\begin{eqnarray}
\mathbb{E}\Delta ^{\prime }C_{i}^{j}\Delta ^{\prime }C_{i^{\prime
}}^{j}&=&q^{j}(1-q^{j})\left( \delta _{i,i^{\prime }}B_{i}^{j}-\frac{%
B_{i}^{j}B_{i^{\prime }}^{j}}{\sum_{i=0}^{2k}B_{i}^{j}}\right)  \nonumber \\
&\cong &q^{j}(1-q^{j})\left( \delta _{i,i^{\prime }}P_{i}^{j}A_{i}-\frac{%
P_{i}^{j}A_{i}P_{i^{\prime }}^{j}A_{i^{\prime }}}{%
\sum_{i=0}^{2k}P_{i}^{j}A_{i}}\right),
\end{eqnarray}
for $j \ne -1$ 
and 
\begin{equation}
\mathbb{E} (\Delta ^{\prime } C^{-1}_{i})^{2} = p_{D}(1-p_{D})A_{i}.
\end{equation}
Other covariances are zero.

Since Bob measures the received pulses with randomly chosen basis, 
the measuring basis coincides with the basis of $\rho_j$ with probability $1/2$ \cite{comment2}.
Therefore, defining the numbers of common basis pulses among $C^{j}_{i}$ by 
$E^{j}_{i}= \frac{1}{2}C^{j}_{i}+ \Delta ^{\prime } E^{j}_{i}$, 
they obey the following binomial distribution
\begin{equation}
P(E_{i}^{j})=\left( \frac{1}{2}\right) ^{C_{i}^{j}}\binom{C_{i}^{j}}{%
E_{i}^{j}},
\end{equation}
for $i=0,1,\ldots,2k$ and $j=-1,\ldots,2k+1$, 
and the nonzero covariances of $\Delta ^{\prime } E^{j}_{i}$ are computed as
\begin{equation}
\mathbb{E} (\Delta ^{\prime } E^{j}_{i})^{2}=
\frac{1}{k+1}C^{j}_{i}\cong \frac{q^{j}P^{j}_{i}A_{i}}{k+1},
\end{equation}
for $j \ne -1$ and
\begin{equation}
\mathbb{E} (\Delta ^{\prime } E^{-1}_{i})^{2}=
\frac{1}{k+1}C^{-1}_{i}\cong \frac{p_{D} A_{i}}{k+1}.
\end{equation}

Now we define the following quantities $F^{j}_{i}$; 
$F^{j}_{i_{0}}$ ($F^{j}_{i_{0}+k}$) denotes the number of check bits with $\times$ ($+$) basis within $E_{i_{0}}-N$ ($N$) bits 
given that the emitted state is $\rho_j$ and $F^{j}_{i}=E^{j}_{i}$ for $i=1,\ldots,k,i \ne i_{0}$.
For $i=i_{0}$ and $i=i_{0}+k$, their distributions are, respectively, given by the following multi-hypergeometric distributions ($j=-1,0,1,\ldots,2k+1$).
\begin{align}
&P(F^{-1}_{i_{0}},F^{0}_{i_{0}},F^{1}_{i_{0}},F^{2}_{i_{0}},
\ldots,F^{k+1}_{i_{0}}) \nonumber \\
=&
\frac{
\binom{E^{-1}_{i_{0}}}{F^{-1}_{i_{0}}}
\binom{E^{0}_{i_{0}}}{F^{0}_{i_{0}}}
\binom{E^{1}_{i_{0}}}{F^{1}_{i_{0}}}
\binom{E^{2}_{i_{0}}}{F^{2}_{i_{0}}}
\cdots 
\binom{E^{k+1}_{i_{0}}}{F^{k+1}_{i_{0}}}
}{\binom{E_{i_{0}}}{E_{i_{0}}-N}},\\
\intertext{and}
& P(F^{-1}_{i_{0}+k},F^{0}_{i_{0}+k},F^{1}_{i_{0}+k},
F^{k+2}_{i_{0}+k},\ldots ,F^{2k+1}_{i_{0}+k}) \nonumber \\
=&
\frac{
\binom{E^{-1}_{i_{0}+k}}{F^{-1}_{i_{0}+k}}
\binom{E^{0}_{i_{0}+k}}{F^{0}_{i_{0}+k}}
\binom{E^{1}_{i_{0}+k}}{F^{1}_{i_{0}+k}}
\binom{E^{k+2}_{i_{0}+k}}{F^{k+2}_{i_{0}+k}}
\cdots
\binom{E^{2k+1}_{i_{0}+k}}{F^{2k+1}_{i_{0}+k}}
}{\binom{E_{i_{0}+k}}{N}},
\end{align}
where
$E_{i}=\sum_{j=-1}^{2k+1} E^{j}_{i}$,
$\sum_{j=-1}^{2k+1} F^{j}_{i_{0}}=E_{i_{0}}-N$,
$\sum_{j=-1}^{2k+1} F^{j}_{i_{0}+k}=N$.
Note that 
$E^{k+2}_{i_{0}}=\cdots =E^{2k+1}_{i_{0}}=E^{2}_{i_{0}+k}=\cdots =E^{k+1}_{i_{0}+k}=0$.
It is easy to see that
\begin{equation}
\mathbb{E} F^{j}_{i_{0}}= \frac{E_{i_{0}}-N}{E_{i_{0}}}E^{j}_{i_{0}},
\end{equation}
and
\begin{equation}
\mathbb{E} F^{j}_{i_{0}+k}= \frac{N}{E_{i_{0}+k}}E^{j}_{i_{0}+k},
\end{equation}
for $j=-1,\ldots,2k+1$.
The nonzero covariances of deviation $\Delta ^{\prime }F_{i}^{j}$ are computed as
\begin{equation}
\mathbb{E} \Delta ^{\prime }F_{i}^{j}\Delta ^{\prime }F_{i}^{j^{\prime }}=
\frac{E_{i}^{j}}{E_{i}}\left( 1-\frac{N}{E_{i}}\right)
\left( \delta ^{j,j^{\prime }}-\frac{NE_{i}^{j^{\prime }}}{E_{i}}\right),
\end{equation}
for $i=i_{0}$ ($j=-1,0,1,2,\ldots,k+1$) and $i=i_{0}+k$ ($j=-1,0,1,k+2,\ldots,2k+1$).

The errors occur among $F_{i}^{j}$ check bits in $\times $ basis with probability $r^{j}$ ($j=1,\ldots,k+1$).
We define $G^{j}_{i}$ as the number of pulses with transmission errors in $\times$ basis
among $F^{j}_{i}$ pulses; $G^{j}_{i}= r^{j} F^{j}_{i}+ \Delta ^{\prime } G^{j}_{i}$, 
which obey the following multi-hypergeometric distribution
\begin{eqnarray}
&&P(G_{1}^{j},\ldots ,G_{k}^{j},G_{i_{0}+k}^{j})  \nonumber \\
&=&\frac{\binom{F_{1}^{j}}{G_{1}^{j}}\cdots \binom{F_{k}^{j}}{G_{k}^{j}}%
\binom{F_{i_{0}+k}^{j}}{G_{i_{0}+k}^{j}}\binom{\sum_{i^{\prime
}=0}^{2k}C_{i^{\prime }}^{j}-\sum_{i^{\prime }=1}^{2k}F_{i^{\prime }}^{j}}{%
r^{j}\sum_{i^{\prime }=0}^{2k}C_{i^{\prime }}^{j}-\sum_{i^{\prime
}=1}^{2k}G_{i^{\prime }}^{j}}}{\binom{\sum_{i^{\prime }=0}^{2k}C_{i^{\prime
}}^{j}}{r^{j}\sum_{i^{\prime }=1}^{2k}C_{i^{\prime }}^{j}}},
\end{eqnarray}
for $i=1,\ldots,k,i_{0}+k$ and $j=1,2,\ldots,k+1$.
Note that $G^{j}_{i}=0$ for $i \ge k+1$ and $i \neq i_{0}+k$ and that the system errors other than the transmission errors are {\em not} counted in the definition of $G^{j}_{i}$.
The nonzero covariances of deviations $\Delta ^{\prime } G^{j}_{i}$ are computed as
\begin{equation}
\mathbb{E} \Delta ^{\prime }G_{i}^{j}\Delta ^{\prime }G_{i^{\prime
}}^{j}=r^{j}(1-r^{j})\left( \delta _{i,i^{\prime }}F_{i}^{j}-\frac{%
F_{i}^{j}F_{i^{\prime }}^{j}}{\sum_{\overline{i}=0}^{2k}C_{\overline{i}}^{j}}%
\right) .
\end{equation}

The number of detected errors in $\times $ basis $H_{i}$ is the sum of several contributions.
The contribution of dark counts (vacuum state) is denoted by $H^{-1}_{i}$ ($H^{0}_{i}$), which is the number of detected errors of $\rho _{-1}$ ($\rho _{0}$) among $F^{-1}_{i}$ ($F^{0}_{i}$) bits.
Since the bits received by Bob are completely independent of the bits sent by Alice for $j=-1$ and $0$, the error probability is $1/2$ so that the probability distributions of random the random variables 
$H^{-1}_{i}$ and  $H^{0}_{i}$ are, respectively, given by
\begin{align}
P(H_{i}^{-1})& =\left( \frac{1}{2}\right) ^{F_{i}^{-1}}\binom{F_{i}^{-1}}{%
H_{i}^{-1}}, \\
\intertext{and}
P(H_{i}^{0})& =\left( \frac{1}{2}\right) ^{F_{i}^{0}}\binom{F_{i}^{0}}{%
H_{i}^{0}}.
\end{align}
For the single photon state, the errors occurred within $G^{1}_{i}$ bits are recovered accidentally by the system errors other than transmission errors with probability $p_{S}$ and that the errors occurred within $F^{1}_{i}-G^{1}_{i}$ by the same cause contribute to the detected errors with probability $p_{S}$.
Therefore, the detected errors of the single photon state are divided into two; 
$H^{1}_{i}$ and ${H^{1}_{i}}^{\prime }$, whose probability distributions are, respectively, given by\begin{align}
P(H^{1}_{i})&=
(1-p_{S})^{H^{1}_{i}}p_{S}^{G^{1}_{i}-H^{1}_{i}}
\binom{G^{1}_{i}}{H^{1}_{i}}, \\
\intertext{and}
P({H^{1}_{i}}^{\prime })&=
p_{S}^{{H^{1}_{i}}^{\prime }}(1-p_{S})^{F^{1}_{i}-G^{1}_{i}-{H^{1}_{i}}^{\prime }}
\binom{F^{1}_{i}-G^{1}_{i}}{{H^{1}_{i}}^{\prime }}.
\end{align}
The random variable $H_{i}$ is then written as
\begin{eqnarray}
H_{i} &=&H_{i}^{-1}+H_{i}^{0}+H_{i}^{1}+H_{i}^{1^{\prime
}}+\sum_{j=2}^{k+1}G_{i}^{j}  \nonumber \\
&=&\frac{1}{2}%
(F_{i}^{-1}+F_{i}^{0})+(1-p_{S})G_{i}^{1}+p_{S}(F_{i}^{1}-G_{i}^{1}) 
\nonumber \\
&&+\sum_{j=2}^{k+1}G_{i}^{j}+\Delta ^{\prime }H_{i},
\end{eqnarray}
where $\Delta ^{\prime } H_{i}$ is the deviation whose nonzero variances are given by
\begin{equation}
\mathbb{E}(\Delta ^{\prime }H_{i})^{2}=\frac{1}{k+1}%
(F_{i}^{-1}+F_{i}^{0})+p_{S}(1-p_{S})F_{i}^{1}.
\end{equation}

The numbers known by Alice and Bob are 
$C_{i}=\sum_{j=-1}^{2k+1} C^{j}_{i}$,
$E_{i}=\sum_{j=-1}^{2k+1} E^{j}_{i}$ for $i=0,1,\ldots,2k$,
and $H_{i}$ for $i=1,\ldots,k$.

\section{Computation of sacrifice key size (Reverse case)} \label{Reverse}

In this section, we give a method to derive the size of sacrifice key of $+$ basis in case of reverse error correction.
According to the central limit theorem with respect to the multi-nomial and 
multi-hypergeometric distributions, 
we can assume safely that all random variables given in Sec.~\ref{Random}
obey normal distributions with 
the averages and the (co)variances given in Sec.~\ref{Random} 
because the number of our samples is sufficiently large.

In the following argument, stochastic variables $B_{i}^{j}$ are fixed.
For now, we fix also 
$x=\frac{q^{k+1}(\mathbf{B})+q^{2k+1}(\mathbf{B})}{\sqrt{2}}$
and $y=r^{k+1}(\mathbf{B})$.
Applying the inequality (\ref{Phase_Error_2}), the quantity 
$P^{\mathcal{P}}_{ph,av,\leftarrow}$ is bounded from above by the expectation of
\[
2^{-\left[ m(\mathcal{D}_{i},\mathcal{D}_{e})-N+F_{i_{0}+k}^{1}\left( 1-%
\overline{h}\left( \frac{G_{i_{0}+k}^{1}}{F_{i_{0}+k}^{1}}\right) \right)
+F_{i_{0}+k}^{-1}\right] _{+}}.
\]
Here, $J^{1}$ and $t$ in (\ref{Phase_Error_2}) are, respectively, given by 
$F_{i_{0}+k}^{1}$ and $G_{i_{0}+k}^{1}$.
The number of pulses detected by dark counts $J^{3}+J^{4}+J^{5}$ in (\ref{Phase_Error_2}) is simply expressed as $F_{i_{0}+k}^{-1}$.
Now, we introduce a new function $\overline{h}_a(x)$ which is a slight modification of $\overline{h}(x)$:
\begin{equation}
\overline{h}_a(x)=
\begin{cases}
\overline{h}(x)& \text{ if } x \ge a, \\
\overline{h}(a)+ \overline{h}^{\prime }(a)(x-a) & \text{ if } x < a.
\end{cases}
\end{equation}
Owing to the convexity of $\overline{h}(x)$, we have 
$\overline{h}(x)\le \overline{h}_{a}(x)$ to obtain
\begin{eqnarray}
&&2^{-\left[ m(\mathcal{D}_{i},\mathcal{D}_{e})-N+F_{i_{0}+k}^{1}\left( 1-%
\overline{h}\left( \frac{G_{i_{0}+k}^{1}}{F_{i_{0}+k}^{1}}\right) \right)
+F_{i_{0}+k}^{-1}\right] _{+}}  \nonumber \\
&\le &2^{-\left[ m(\mathcal{D}_{i},\mathcal{D}_{e})-N+F_{i_{0}+k}^{1}\left(
1-\overline{h}_{a}\left( \frac{G_{i_{0}+k}^{1}}{F_{i_{0}+k}^{1}}\right)
\right) +F_{i_{0}+k}^{-1}\right] _{+}},
\end{eqnarray}
which is used for an upper bound of the quantity $P^{\mathcal{P}}_{ph,av,\leftarrow}$.
Here, we estimate
\begin{equation}
\Theta (\mathcal{D}_{i},\mathcal{D}_{e}):=N-F_{i_{0}+k}^{1}\left[ 1-%
\overline{h}_{a}\left( \frac{G_{i_{0}+k}^{1}}{F_{i_{0}+k}^{1}}\right)
\right] -F_{i_{0}+k}^{-1},
\end{equation}
by the estimator \cite{comment3}
\begin{equation}
\hat{\Theta}_{x,y}(\mathcal{D}_{i},\mathcal{D}_{e}):=N-\frac{NA_{i_{0}+k}}{%
C_{i_{0}+k}}\left\{ \hat{q}_{x}^{1}P_{i_{0}+k}^{1}\left[ 1-\overline{h}_{a}(%
\hat{r}_{x,y}^{1})\right] +p_{D}\right\} .
\end{equation}
The deviation $\Theta (\mathcal{D}_{i},\mathcal{D}_{e})-\hat{\Theta }_{x,y}(\mathcal{D}_{i},%
\mathcal{D}_{e})$ is divided into two stochastic variables, 
$\Delta \Theta _{1}$ and $\Delta \Theta _{2}$;
\begin{equation}
\Theta (\mathcal{D}_{i},\mathcal{D}_{e})-\hat{\Theta }_{x,y}(\mathcal{D}_{i},%
\mathcal{D}_{e}) =\Delta \Theta _{1}+\Delta \Theta _{2},
\end{equation}
where
\begin{eqnarray}
\Delta \Theta _{1}&:=&-F_{i_{0}+k}^{1}\left[ 1-\overline{h}_{a}\left( \frac{%
G_{i_{0}+k}^{1}}{F_{i_{0}+k}^{1}}\right) \right] -F_{i_{0}+k}^{-1}  \nonumber
\\
&&+\frac{N\left\{ \hat{q}_{x}^{1}B_{i_{0}+k}^{1}\left[ 1-\overline{h}_{a}(%
\hat{r}_{x,y}^{1})\right] +A_{i_{0}+k}p_{D}\right\} }{C_{i_{0}+k}},
\end{eqnarray}
and
\begin{equation}
\Delta \Theta _{2}:=\frac{N\hat{q}^{1}_{x}(A_{i_{0}+k}P_{i_{0}+k}^{1}-B_{i_{0}+k}^{1})%
\left[ 1-\overline{h}_{a}(\hat{r}^{1}_{x,y})\right] }{C_{i_{0}+k}}.
\label{Theta_2}
\end{equation}

Now, let us apply the Gaussian approximation to the variables 
$\Delta \Theta _{1}$ and $\Delta \Theta _{2}$.
Since the mean of $\Delta \Theta _{1}$ is zero, 
\begin{equation}
\sqrt{v_{i_{0;x,y}}(\mathbf{q},\mathbf{r},B_{i}^{j},\mathbf{A},\pmb{\mu})}
\Phi ^{-1}(2^{-\delta _{1}})\ge \Delta \Theta _{1},
\end{equation}
with probability $\geq 1-2^{-\delta_{1}}$.
Here, 
$v_{i_{0;x,y}}(\mathbf{q},\mathbf{r},B_{i}^{j},\mathbf{A},\pmb{\mu})$ 
denotes the variance of $\Delta \Theta _{1}$,
and
\begin{equation}
\Phi(x)=\frac{1}{\sqrt{2\pi}}
\int_{-\infty}^{-x} e^{-\frac{y^{2}}{2}}dy
\end{equation}
is the probabilistic distribution of the standard normal distribution.
In Eq.~(\ref{Theta_2}), 
$A_{i_{0}+k} P^{1}_{i_{0}+k} -B^{1}_{i_{0}+k}$ is a stochastic variable with the mean zero and the variance $A_{i_{0}+k} P^{1}_{i_{0}+k}(1-P^{1}_{i_{0}+k})$.
It follows that
\begin{equation}
\frac{N\hat{q}_{x}^{1}\left[ 1-\overline{h}_{a}(\hat{r}_{x,y}^{1})\right] }{%
C_{i_{0}+k}}\sqrt{A_{i_{0}+k}P_{i_{0}+k}^{1}(1-P_{i_{0}+k}^{1})}\Phi
^{-1}(2^{-\delta _{2}})\ge \Delta \Theta _{2},
\end{equation}
with probability $\geq 1-2^{-\delta _{2}}$.
Consequently, taking the size of privacy amplification $m(\mathcal{D}_{i},\mathcal{D}_{e})$ in (\ref{Phase_Error_2}) as
\begin{eqnarray}
&&m(\mathcal{D}_{i},\mathcal{D}_{e}) =\hat{\Theta}_{x,y}(\mathcal{D}_{i},\mathcal{D%
}_{e})+\sqrt{v_{i_{0;x,y}}(\mathbf{q},\mathbf{r},B_{i}^{j},\mathbf{A},%
\pmb{\mu})}\Phi ^{-1}(2^{-\delta _{1}})  \nonumber \\
&&+\frac{N\hat{q}_{x}^{1}\left[ 1-\overline{h}_{a}(\hat{r}_{x,y}^{1})\right] 
}{C_{i_{0}+k}}\sqrt{A_{i_{0}+k}P_{i_{0}+k}^{1}(1-P_{i_{0}+k}^{1})}\Phi
^{-1}(2^{-\delta _{2}}) \nonumber \\
&&+\delta _{3},
\label{m}
\end{eqnarray}
we have
\begin{equation}
P^{\mathcal{P}}_{ph,av,\leftarrow}\le 2^{-\delta_{1}}+2^{-\delta_{2}}+2^{-\delta_{3}}.
\label{Phase_Error_3}
\end{equation}
The quantity $m(\mathcal{D}_{i},\mathcal{D}_{e})$ [Eq.~(\ref{m})] depends on $\mathbf{q}$, $\mathbf{r}$, and $B_{i}^{j}$, which can be approximated by 
$\mathbf{q}^{ML}_{x,y,(B^{j}_{i})}$, $\mathbf{r}^{ML}_{x,y,(B^{j}_{i})}$, and $A_{i} P^{j}_{i}$, respectively 
so that $m(\mathcal{D}_{i},\mathcal{D}_{e})$ can be written in terms of observed quantities.
From above observation, we define the size of privacy amplification for fixed $x$ and $y$ as
\begin{eqnarray}
&&m_{i_{0};x,y}(\mathcal{D}_{i},\mathcal{D}_{e})  \nonumber \\
:= &&\hat{\Theta}_{x,y}(\mathcal{D}_{i},\mathcal{D}_{e})  \nonumber \\
&&+\sqrt{v_{i_{0;x,y}}(\mathbf{q}_{x,y,(A_{i}P_{i}^{j})}^{ML},\mathbf{r}%
_{x,y,(A_{i}P_{i}^{j})}^{ML},A_{i} P_{i}^{j},\mathbf{A},\pmb{\mu})}\Phi
^{-1}(2^{-\delta _{1}}) \nonumber \\
&&+\frac{N\hat{q}_{x}^{1}\left[ 1-\overline{h}_{a}(\hat{r}_{x,y}^{1})\right] 
}{C_{i_{0}+k}}\sqrt{A_{i_{0}+k}P_{i_{0}+k}^{1}(1-P_{i_{0}+k}^{1})}\Phi
^{-1}(2^{-\delta _{2}})  \nonumber \\
&&+\delta _{3},
\label{PA}
\end{eqnarray}
which satisfies
\begin{equation}
m_{i_{0};x,y}(\mathcal{D}_{i},\mathcal{D}_{e})\ge 
\Theta (\mathcal{D}_{i},\mathcal{D}_{e})+\delta _{3}
\end{equation}
with probability $1- 2^{-\delta _{1}}- 2^{-\delta _{2}}$.
It should be noted that the linear estimators $\hat{q}_{x}^{1}$ and $\hat{r}_{x,y}^{1}$ are used in the first term of the right-hand side of Eq.~(\ref{PA}), while the maximally likelihood estimators are used in the second and third terms.
This is because if linear estimators were used in the second and third terms in the right-hand side of Eq.~(\ref{PA}), these terms would not well-defined since the linear estimators do not necessarily satisfy 
$0\leq \hat{q}_{x}^{j}\leq 1-p_{D}$ and $0\leq \hat{r}_{x,y}^{j}\leq 1$.

Now, we take the worst case and define the size of privacy amplification $m_{i_{0}}(\mathcal{D}_{i},\mathcal{D}_{e})$ as
\begin{equation}
m_{i_{0}}(\mathcal{D}_{i},\mathcal{D}_{e}):=
\max_{0\le x\le \sqrt{2}(1-p_{D}),0 \le y\le 1}
m_{i_{0};x,y}(\mathcal{D}_{i},\mathcal{D}_{e}),
\end{equation}
then (\ref{Phase_Error_3}) holds for arbitrary $x$ and $y$ \cite{comment4}.

The size of privacy amplification for $\times$ basis is given by
$m_{i_{0}}({\tilde{\mathbf{A}}},\pmb{\mu},\tilde{p}_{S},p_{D};
{\tilde{\mathbf{C}}},{\tilde{\mathbf{E}}},
{\tilde{\mathbf{H}}})$
where 
$\tilde{A}_{0}=A_{0} $, 
$\tilde{A}_{i}=A_{i+k} $,
$\tilde{A}_{i+k}=A_{i} $,
$\tilde{C}_{0}=C_{0} $, 
$\tilde{C}_{i}=C_{i+k} $,
$\tilde{C}_{i+k}=C_{i} $,
$\tilde{E}_{0}=E_{0} $, 
$\tilde{E}_{i}=E_{i+k} $,
$\tilde{E}_{i+k}=E_{i} $,
$\tilde{H}_{i}=H_{i+k} $, and 
$\tilde{H}_{i+k}=H_{i} $
for $i=1,\ldots,k$.

In the course of actual numerical computations, 
it often happens that $x$ or $y$ moves away from its defining region.
To circumvent this difficulty, we use $\overline{h}_a(z)$ instead of $\overline{h}(z)$, which enables us to extend the accessible region of $z$ to $(- \infty,\infty)$ and further to avoid the divergence of the derivative of $\overline{h}$.
When $a$ is small enough, $v_{i_{0;x,y}}(\mathbf{q},\mathbf{r},B_{i}^{j},\mathbf{A},\pmb{\mu})$ takes on a large value, while $a$ is large, 
$\hat{\Theta}_{x,y}(\mathcal{D}_{i},\mathcal{D}_{e})$ takes on a large value in turn.
The parameter $a$ must be properly chosen taking into account such a trade-off behavior.
The detail is shown in Appendix~\ref{sec:param_a}.

Now, let us go back to Eq.~(\ref{Information_Inequality_1}).
To ensure
\begin{equation}
I^{\mathcal{P}}_{E,av,\leftarrow}\le 2^{-\delta},
\label{Information_Inequality_2}
\end{equation}
it is sufficient to choose 
$\delta _{1}=\delta +\delta ^{\prime }+1$ and 
$\delta _{2}=\delta _{3}=\delta+\delta ^{\prime }+2$.
Here, $\delta^{\prime }$ is $\left\lceil \log_{2} \overline{N}\right\rceil $.
Since $\delta +\delta^{\prime }\ll \overline{N}$, 
$\delta+\delta^{\prime }+1+\overline{N}\lesssim 2^{\delta^{\prime }}$.
Thus,
\begin{eqnarray}
 I^{\mathcal{P}}_{E,av,\leftarrow}
  &\le& P^{\mathcal{P}}_{ph,av,\leftarrow} (1+\overline{N}-\log
  P^{\mathcal{P}}_{ph,av,\leftarrow}) \notag \\
 &\le& 2^{-\delta-\delta'}(1+\overline{N}+\delta+\delta') \notag \\
 &\le& 2^{-\delta-\delta'}\cdot 2^{\delta'} = 2^{-\delta}.
\end{eqnarray}

In parallel with the above argument based on the Gaussian approximation, 
the large deviation type evaluation is also possible.
By Cram\'er's theorem \cite{Cramer}, 
$\mathrm{Prob}\{|X-\mathbb{E} (X)|> c N\}$
goes to zero exponentially
when $X$ obeys the $N$ trials of a multinomial distribution 
or a multinomial hypergeometric distribution for an arbitrary constant $c>0$.
Hence, the probability satisfying the inequalities
\begin{equation}
\Delta \Theta _{2} < c_{1} N,
\label{Theta_2_Inequality}
\end{equation}
$\Delta F^1_{i_0+k} = F^1_{i_0+k}-\mathbb{E}(F^1_{i_0+k})< c_{2} N$, 
$\Delta G^1_{i_0+k} = G^1_{i_0+k}-\mathbb{E}(G^1_{i_0+k})< c_{3} N$, and 
$\Delta F^{-1}_{i_0+k} = F^{-1}_{i_0+k}-\mathbb{E}(F^{-1}_{i_0+k})< c_{4} N$ 
goes to zero exponentially for arbitrary constants $c_{1},c_{2},c_{3},c_{4}>0$.
For an arbitrary constant $c_{5}>0$,
we choose $c_{2},c_{3},c_{4}>0$ such that
if $\Delta F_{i_{0}+k}^{1}<Nc_{2}$, 
$\Delta G_{i_{0}+k}^{1}<Nc_{3},\Delta$, and
$\Delta F_{i_{0}+k}^{-1}<Nc_{4}$, then 
\begin{equation}
\Delta \Theta _{1}<c_{5}N
\label{Theta_1_Inequality}
\end{equation}
is always satisfied.
Thus, the probability satisfying (\ref{Theta_1_Inequality}) goes to zero exponentially.
Now, we denote the exponential upper bounds of 
(\ref{Theta_2_Inequality}) and (\ref{Theta_1_Inequality}) 
by $2^{-d_{1} N}$ and $2^{-d_{2} N}$, respectively, 
and choose the size of sacrifice bits $m_{i_0:x,y}(\mathcal{D}_{i},\mathcal{D}_{e})$ 
as 
$\mathbb{E}\Theta +(c_{1}+c_{5}+c_{6})N$ 
for a constant $c_6 >0$.
Then, applying (\ref{Phase_Error_2}), we obtain an exponential upper bound:
\begin{equation}
P^{\mathcal{P}}_{ph,av,\leftarrow}\le 
2^{-d_{1} N}+2^{-d_{2} N}+2^{-c_6 N}.
\end{equation}

However, as is known in mathematical statistics, 
the above exponential evaluation does not yield a tighter upper bound of 
$\mathrm{Prob}\{|X-\mathbb{E} (X)|> c N\}$ 
as the Gaussian approximation does.
Hence, in the finite-length code, 
the Gaussian approximation gives a tighter upper bound of $P^{\mathcal{P}}_{ph,av,\leftarrow}$
than the above exponential upper bound.
This is the reason why we have employed the Gaussian approximation.

\section{Computation of $v_{i_{0;x,y}}(\mathbf{q},\mathbf{r},B_{i}^{j},\mathbf{A},\pmb{\mu})$}
\label{v}

The variance $v_{i_{0;x,y}}(\mathbf{q},\mathbf{r},B_{i}^{j},\mathbf{A},\pmb{\mu})$ 
for a given stochastic variable $B^{j}_{i}$ and under constraints 
$\frac{q^{k+1}+q^{2k+1}}{\sqrt{2}}=x$ and $r^{k+1}=y$
is given by
\begin{equation}
v_{i_{0;x,y}}(\mathbf{q},\mathbf{r},B_{i}^{j},\mathbf{A},\pmb{\mu})=
 \mathbb{E}\Delta \Theta _{1}^{2},
\label{Variance}
\end{equation}
where
\begin{align}
& \Delta \Theta _{1}  \nonumber \\
\cong & -\left[ 1-\overline{h}_{a}(r^{1})\right] \Delta F_{i_{0}+k}^{1}+%
\overline{h}_{a}^{\prime }(r^{1})\Delta ^{\prime }G_{i_{0}+k}^{1}-\Delta
F_{i_{0}+k}^{-1}  \nonumber \\
& -\frac{N\left\{ q^{1}B_{i_{0}+k}^{1}\left[ 1-\overline{h}%
_{a}(r^{1})\right] +A_{i_{0}+k}p_{D}\right\} }{(\mathbb{E}C_{i_{0}+k})^{2}}%
\Delta C_{i_{0}+k}  \nonumber \\
& +\frac{NB_{i_{0}+k}^{1}\left[ 1-\overline{h}_{a}(r^{1})\right] }{\mathbb{E}%
C_{i_{0}+k}}\Delta \hat{q}_{x}^{1}-\frac{N}{\mathbb{E}C_{i_{0}+k}}%
q^{1}B_{i_{0}+k}^{1}\overline{h}_{a}^{\prime }(r^{1})\Delta \hat{r}%
_{x,y}^{1},
\label{Deviation}
\end{align}
with $\Delta X = X- \mathbb{E} X$ in the right-hand side of Eq.~(\ref{Deviation}).
Here, stochastic variables, 
$\Delta F^{1}_{i_{0}+k}$, $\Delta F^{-1}_{i_{0}+k}$, 
$\Delta \hat{q}^{1}_{x}$, and $\Delta \hat{r}^{1} _{x,y}$ are computed as
\begin{eqnarray}
\Delta F_{i_{0}+k}^{j} &\cong &\frac{2N}{\mathbb{E}C_{i_{0}+k}}\Delta
E_{i_{0}+k}^{j}-\frac{2Nq^{j}B_{i_{0}+k}^{j}}{(\mathbb{E}C_{i_{0}+k})^{2}}%
\sum_{j^{\prime }=-1}^{2k+1}\Delta E_{i_{0}+k}^{j^{\prime }}
%\newline
\nonumber \\
&&+\Delta ^{\prime }F_{i_{0}+k}^{j},
\label{Deviation_F}
\end{eqnarray}
\begin{equation}
\Delta \hat{q}_{x}^{1}=\sum_{i=1}^{2k}(\overline{P}_{2k\times
2k}^{-1})_{1}^{i}\left( \frac{\Delta C_{i}}{A_{i}}-\overline{P}_{i}^{0}\frac{%
\Delta C_{0}}{A_{i}}\right),
\end{equation}
and
\begin{widetext}
\begin{eqnarray}
\Delta \hat{r}_{x,y}^{1} &\cong &-\frac{1}{(1-2p_{S})(\hat{q}_{x}^{1})^{2}}%
\sum_{i=1}^{k}(P_{k\times k}^{-1})_{1}^{i}\left\{ \mathbb{E}(s_{i}p_{i})-%
\frac{1}{2}\left[ P_{i}^{0}\left( \frac{\mathbb{E}C_{0}}{A_{0}}-p_{D}\right)
+p_{D}\right] -P_{i}^{k+1}q^{k+1}y\right\} \Delta \hat{q}_{x}^{1}
\nonumber \\
&&+\frac{1}{(1-2p_{S})\hat{q}_{x}^{1}}\sum_{i=1}^{k}(P_{k\times
k}^{-1})_{1}^{i}\left( \Delta (s_{i}p_{i})-\frac{1}{2}P_{i}^{0}\frac{\Delta
C_{0}}{A_{0}}-P_{i}^{k+1}y\Delta \hat{q}_{x}^{k+1}\right).
\label{Deviation_r}
\end{eqnarray}
\end{widetext}
In Eq.~(\ref{Deviation_r}), $\Delta \hat{q}^{k+1}_{x} $ and $\Delta (s_{i} p_{i})$ are given by
\begin{equation}
\Delta \hat{q}_{x}^{k+1}=\frac{1}{\sqrt{2}}\sum_{i=1}^{2k}(\overline{P}%
_{2k\times 2k}^{-1})_{k+1}^{i}\left( \frac{\Delta C_{i}}{A_{i}}-\overline{P}%
_{i}^{0}\frac{\Delta C_{0}}{A_{i}}\right),
\end{equation}
\begin{equation}
\Delta (s_{i} p_{i})
\cong 
\frac{2 }{A_{i}} \Delta H_{i}
-\frac{2\mathbb{E} H_{i}}{A_{i} \mathbb{E} E_{i}} \Delta ^{\prime } E_{i},
\label{Deviation_sp_1}
\end{equation}
for $i=1,\ldots,k; i \ne i_{0}$ and
\begin{eqnarray}
\Delta (s_{i_{0}}p_{i_{0}}) &\cong &\frac{\mathbb{E}C_{i_{0}}}{A_{i_{0}}(%
\mathbb{E}E_{i_{0}}-N)}\Delta H_{i_{0}}-\frac{\mathbb{E}C_{i_{0}}\mathbb{E}%
H_{i_{0}}}{A_{i_{0}}(\mathbb{E}E_{i_{0}}-N)^{2}}\Delta ^{\prime }E_{i_{0}} \nonumber \\
&&-\frac{N\mathbb{E} H_{i_{0}}}{A_{i_{0}}(\mathbb{E}E_{i_{0}}-N)^{2}}\Delta
C_{i_{0}}.
\label{Deviation_sp_2}
\end{eqnarray}
Deviations $\Delta H_{i}$ in Eqs.~(\ref{Deviation_sp_1}) and (\ref{Deviation_sp_2}) are computed as
\begin{eqnarray}
\Delta H_{i} &=&\sum_{j=-1}^{k+1}\frac{\tilde{r}^{j}}{2}\Delta
C_{i}^{j}+\sum_{j=-1}^{k+1}\tilde{r}^{j}\Delta ^{\prime
}E_{i}^{j}+(1-2p_{S})\Delta ^{\prime }G_{i}^{1} \nonumber \\
&&+\sum_{j=2}^{k+1}\Delta ^{\prime }G_{i}^{j}+\Delta ^{\prime }H_{i},
\end{eqnarray}
for $i = 1, \ldots,k; i \ne i_{0}$.
and
\begin{eqnarray}
\Delta H_{i_{0}} &=&\sum_{j=-1}^{k+1}\tilde{r}^{j}\left[ \left( 1-\frac{2N}{%
\mathbb{E}C_{i_{0}}}\right) \Delta E_{i_{0}}^{j}\right.  \nonumber \\
&&\left. +\frac{2N\mathbb{E}C_{i_{0}}^{j}}{(\mathbb{E}C_{i_{0}})^{2}}%
\sum_{j^{^{\prime }}=-1}^{2k+1}\Delta E_{i_{0}}^{j^{^{\prime }}}+\Delta
^{\prime }F_{i_{0}}^{j}\right]  \nonumber \\
&&+(1-2p_{S})\Delta ^{\prime }G_{i_{0}}^{1}+\sum_{j=2}^{k+1}\Delta ^{\prime
}G_{i_{0}}^{j}+\Delta ^{\prime }H_{i_{0}},
\label{Deviation_H}
\end{eqnarray}
where 
$\tilde{r}^{-1}=\tilde{r}^{0}=\frac{1}{2}$, 
$\tilde{r}^{1}=r^{1} +(1-2 r^{1})p_{S}$, and 
$\tilde{r}^{j}=r^{j}$ for $j \ge 2$.
Note that 
$\Delta E_{i}^{j}=\frac{1}{2}\Delta ^{^{\prime }}C_{i}^{j}+\Delta ^{^{\prime
}}E_{i}^{j}$ 
in Eqs.~(\ref{Deviation_F}) and (\ref{Deviation_H}).
Equation~(\ref{Variance}) with subsequent equations in this section and (co)variances given in Sec.~\ref{Random} yields the explicit form of $v_{i_{0;x,y}}$.

If $\mathbf{q}$ and $\mathbf{r}$ coincide with the respective values estimated from the observed quantities 
$\mathcal{D}_{e}=(\mathbf{C}, \mathbf{E},\mathbf{H})$, 
$\mathbb{E} C_{i}$, $\mathbb{E} E_{i_{0}}$, and $\mathbb{E} H_{i}$ in Eqs.~(\ref{Deviation}), (\ref{Deviation_F}), (\ref{Deviation_sp_1}), and (\ref{Deviation_sp_2}) are, respectively, equal to the observed quantities, $C_{i}$, $E_{i_{0}}$, and $H_{i}$.

\section{Numerical Analysis} \label{Numerical}

In this section we show the results of numerical simulation
to reveal performances of our protocol
and to know the optimal values of parameters such as intensities
in our protocol.
For numerical simulation, we use $k+1=4$ different intensities
including vacuum because $k=3$ is
required for at least good estimation of the probability
that multi-photon is actually sent by Alice~\cite{Hayashi_Asymptotic}.

\subsection{Parameters}

Let $a_1\ \mbox{(db/km)}$ be the fiber loss,
$a_0\ \mbox{(db)}$ be the receiver loss,
$\eta_{det}$ be the efficiency of the detector, and
$L\ (\mbox{km})$ be the transmission distance.
A detection probability $p$ of the pulse $\mu_i$ can be
represented by
\begin{equation}
 p=1-e^{-\alpha \mu_i}+p_0,
\end{equation}
where
\begin{equation}
 \alpha=\eta_{det} \cdot 10^{-\frac{a_1 L + a_0}{10}}.
\end{equation}
Listed in Table~\ref{table:parameters} are parameters
used for our numerical simulation, all of which
are experimental values in the long distance experiment~\cite{KNHTKN},
and $a_1$ is the lowest loss value in commercially available optical
fibers~\cite{optical_fiber}.
We assume that the detection probability $p_0$ of vacuum state
equals to dark count rate $p_D$.
\begin{table}
 \caption{Parameters for numerical simulation.}
 \label{table:parameters}
 \begin{center}
  \begin{tabular}{ccc} \hline \hline
   $a_1$ (db/km) & $a_0$ (db) & $\eta_{det}$ \\
   0.17 & 5.0 & 0.1 \\ \hline
   $p_{D}$ & $p_{S}, \tilde{p}_{S}$ & $p_0$ \\
   $4.0 \times 10^{-7}$ & $3\%$ & $4.0 \times 10^{-7}$ \\ \hline \hline
  \end{tabular}
 \end{center}
\end{table}
In this setting, the detection probability of the pulse $\mu=0.5$
at $L=20.0 (100.0) \ \mbox{(km)}$ is $7.2\times 10^{-3}(
3.2\times 10^{-3})$ and
the error probability is $3.00\%(3.06\%)$, respectively.

We also fix the security parameter $\delta$ in order to satisfy
the average of Eve's information $I^{\mathcal{P}}_{E,av,\leftarrow}\le 2^{-9}$
in Eq.~(\ref{Information_Inequality_2}).
It is sufficient to set 
$\delta_{1}=\delta_{3}=9+[\log_{2} (\overline{N})]+2$, and 
$\delta_{2}=9+[\log_{2} (\overline{N})]+1$.
%$\Phi^{-1}(2^{-\delta_{1}})\cong  6$.

\begin{figure}
 \begin{center}
  \includegraphics[width=0.49\textwidth]{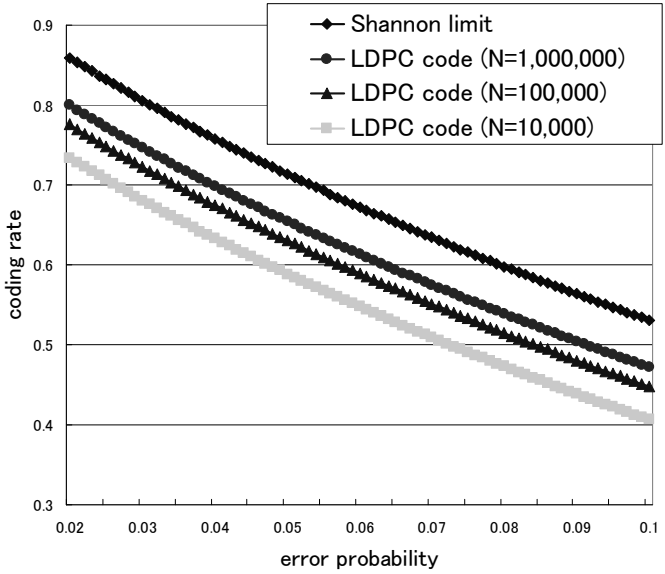}
 \end{center}
 \caption{Coding rates of the LDPC codes
 used for numerical simulation.}
 \label{fig:coding_rate}
\end{figure}

For the error correction with the finite code length,
we use the LDPC (Low Density Parity Check) code~\cite{Gallager,MacKay}
which is known to be one of the best classical error correcting codes
and the performance can asymptotically achieve the Shannon limit.
Figure~\ref{fig:coding_rate} shows coding rates of our LDPC codes.
The coding rate with $N=1.0 \times 10^6$ is about $0.75$
when the error probability is $3\%$.

Suppose that intensities $\mu_i$ and sending probabilities $\tilde{p}_i$
take discrete values $0.05, 0.10, \dots, 1.00$
 for our numerical simulation
because computation of $m_{i_{0};x,y}(\mathcal{D}_{i},\mathcal{D}_{e})$~(\ref{PA})
needs a constrained non-linear optimization
and it is therefore rather hard and time-consuming task.

\subsection{Performance}

\begin{figure}[t]
 \begin{center}
  \includegraphics[width=0.49\textwidth]{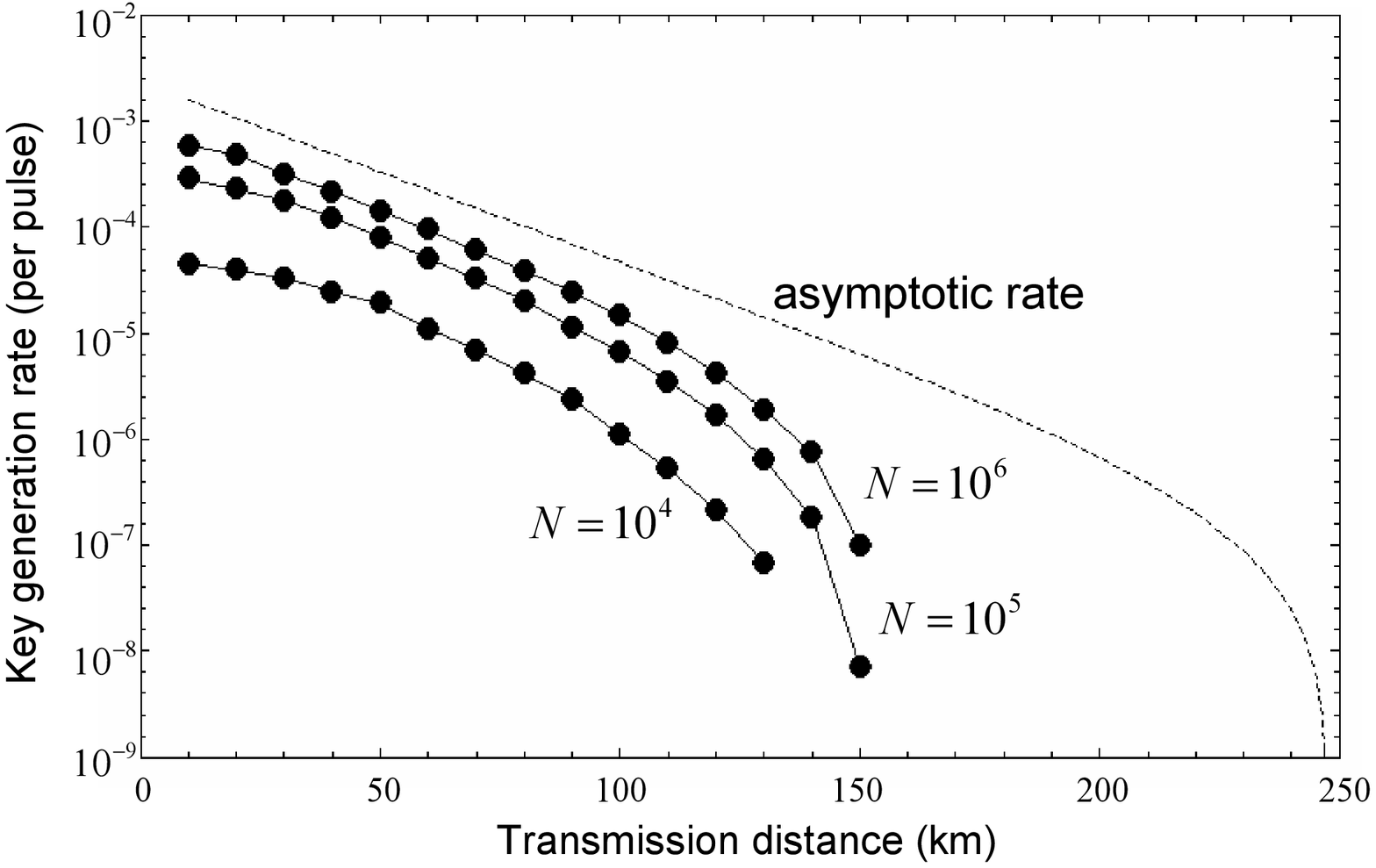}
 \end{center}
 \caption{Performance of our decoy state QKD.
 The secure secret key can be shared up to 150 km when $N=1.0 \times 10^{6}$.}
 \label{fig:rate_distance}
%\end{figure}
%\begin{figure}
 \begin{center}
  \includegraphics[width=0.49\textwidth]{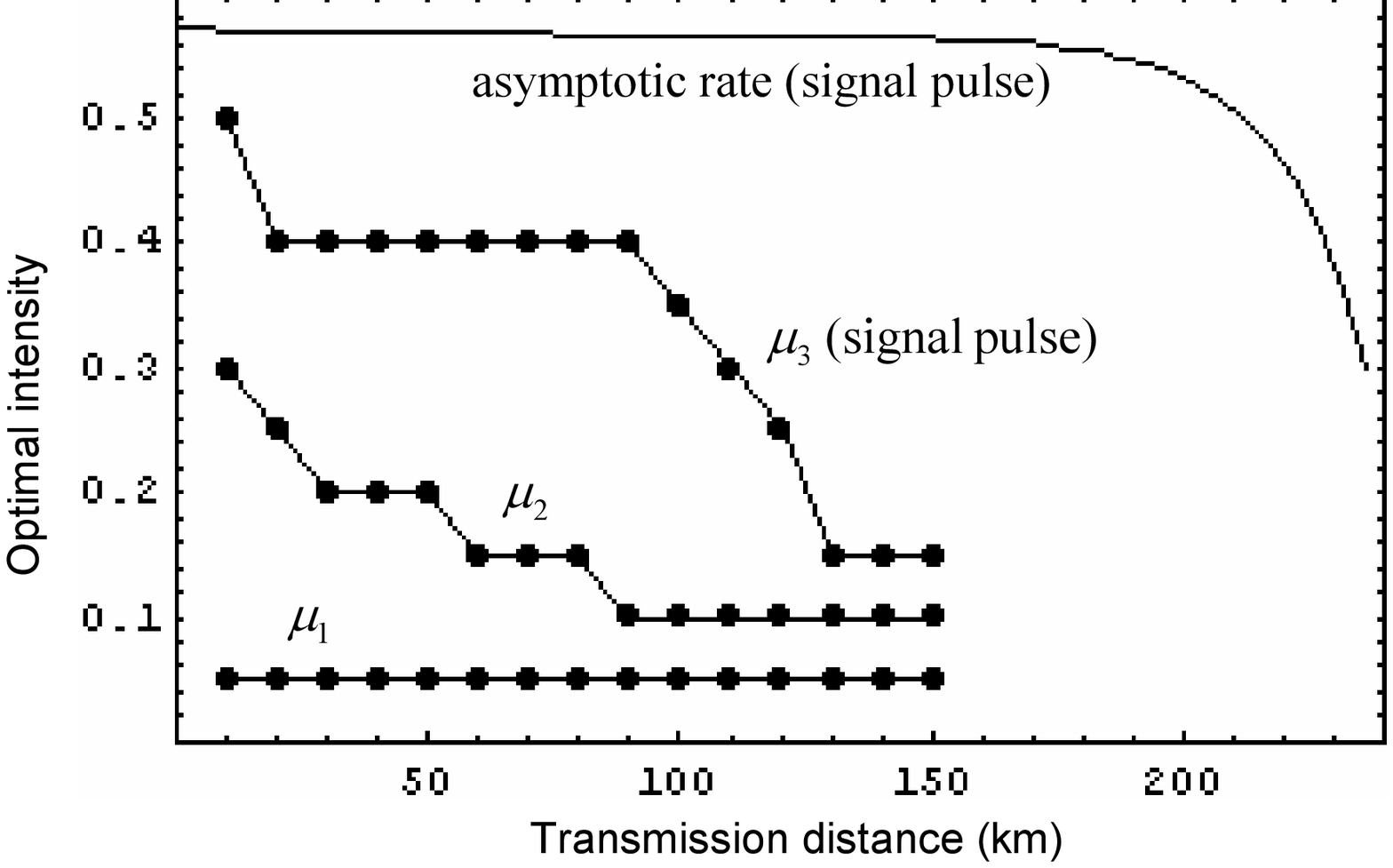}
 \end{center}
 \caption{Optimal intensities with $N=1.0 \times 10^6$ and
 the asymptotic case corresponding to the key generation rates
 in Figure~\ref{fig:rate_distance}.
 %The optimal intensity with $1.0\times 10^6$ at 20 km is 0.45.
 }
 \label{fig:optimal_intensity}
\end{figure}

We first show key generation rates of our protocol
with respect to the transmission distance.
Figure~\ref{fig:rate_distance} shows the optimal key generation rates per pulse sent
by Alice when the code length $N$ equals to $1.0\times 10^4$,
$1.0\times 10^5$, and $1.0\times 10^6$, keeping
the average of Eve's information less than $2^{-9}$.
For comparison, the asymptotic rate is also added,
which is calculated from the asymptotic rate formula
with three different intensities including vacuum~\cite{Hayashi_Asymptotic}.
Our decoy state QKD enables Alice and Bob to share the final
secret key securely up to 150 (km)
while the secure secret key can be shared up to 250 (km) in the asymptotic case.
Although there is a huge gap between the maximum
transmission distances of the finite and asymptotic case,
such small key generation rate with the finite code length,
in other words the large number of sacrifice bits,
is needed for keeping the average of Eve's information
less than $2^{-9}$ incorporating the statistical fluctuations.
The main reason why the more key generation rate is obtained
with larger code size is that Eve's information can be
estimated better with large size statistics,
such that the larger number of sampling given by the check bits
can be used for estimation
as well as the finite error correcting coding rate
in Figure~\ref{fig:coding_rate}.

We then show the optimal intensities
corresponding to the rates in Figure~\ref{fig:rate_distance}.
The three optimal different intensities $\mu_i$ with $N=1.0 \times 10^6$
as well as the signal intensity in the asymptotic case
are shown in Figure~\ref{fig:optimal_intensity}.
The optimal signal intensities
decrease as the transmission distance is longer.
%are smaller for the longer transmission distance.
This tendency is easy to understood
because the probability that the multi-photon is emitted
cannot be negligible for estimation of the quantum channel
at the long transmission distance.
At the maximum transmission distance 150 (km),
The optimal intensity $\mu_3$ becomes $0.15$
which is the minimum value of the signal intensity
available for our numerical simulation
because intensities $\mu_i$ take discrete values by 0.05
and $0 < \mu_1 < \mu_2 < \mu_3$.
It is quite smaller than the intensity $0.3$ of the asymptotic case
at the maximal transmission distance 250 (km).
This is because
considering statistical fluctuations including estimation errors
causes the worse estimations of
$m_{i_{0};x,y}(\mathcal{D}_{i},\mathcal{D}_{e})$
and such random variables as $\mathbf{J}$
than the true values without estimation errors.
It turns out that the probability that the multi-photon is sent
including statistical fluctuations becomes more dominant
factor in our protocol.

\begin{figure}
 \begin{center}
  \includegraphics[width=0.49\textwidth]{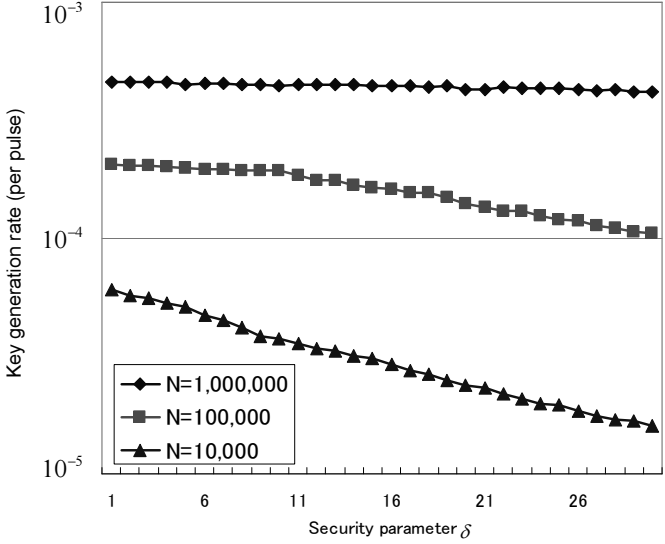}
 \end{center}
 \caption{Key generation rates with respect to the security parameter
 $\delta$ when the code length $N=10^4, 10^5,$ and $10^6$.}
 \label{fig:rate_security}
\end{figure}
We finally examine the performance of our protocol
with respect to the security parameter $\delta$.
Figure~\ref{fig:rate_security} shows the key generation rates
at the transmission distance 20 (km).
The rates when $N=1.0 \times 10^6$
hardly decrease at all
while the rates when $N=1.0 \times 10^4$ and $1.0 \times 10^5$
become smaller as $\delta$ is bigger.
This is because the larger number of sampling when $N$ is large enough
makes the variance $v_{i_{0;x,y}}$ in Eq~(\ref{PA}) smaller.
Therefore
the size of sacrifice bits
$m_{i_{0};x,y}(\mathcal{D}_{i},\mathcal{D}_{e})$
is affected little by $\delta_1$, that is $\delta$,
and the final secret key can be shared
as securely as you want when $N=1.0 \times 10^6$.

\section{Conclusions and future perspectives} \label{Conclusions}

In conclusion, we have derived a formula for the size the final secret key in the finite code length decoy state QKD with arbitrary number of decoy states of different intensities incorporating the finite statistics~\cite{comment5}.

We have utilized the central limit theorem and thereby neglected the higher-order terms in the formula obtained here.
Furthermore, there is room to improve the derivation of the size of privacy amplification.
These points will be further pursued in the future.
Finally, we have assumed that the error probability of the signal generation at the sending port is unknown.
If this is not the case, the generation rate of the final secret key would be improved \cite{RGK}.
Such an improvement of arguments in Sec.~\ref{Reverse} would be also one of the future problems.

\appendix

\section{Choice of parameter $a$} \label{sec:param_a}

In Sec.~\ref{Reverse} we have used $\overline{h}_a(x)$ instead of $\overline{h}(x)$ to circumvent the singularity of $\overline{h}(x)$ when $x\rightarrow 0$.
If $x $ is so close to $0$ that 
$\overline{h}_{a}^{\prime }(x)\gg $ $\overline{h}_{a}(x)$, 
the leading term of $\Delta \Theta _{1}$ is well approximated by
\begin{equation*}
\overline{h}_{a}^{\prime }(\hat{r}^{1}_{x,y})\left( \Delta ^{\prime }G_{i_{0}+k}^{1}-%
\frac{N}{\mathbb{E} C_{i_{0}+k}}q^{1}B_{i_{0}+k}^{1}\Delta \hat{r}_{x,y}^{1}\right).
\end{equation*}
Hence, denoting the variance of
\begin{equation*}
\Delta ^{\prime } G^{1}_{i_{0}+k}
-\frac{N}{\mathbb{E} C_{i_{0}+k}}q^{1} B^{1}_{i_{0}+k} \Delta \hat{r}^{1} _{x,y}
\end{equation*}
by $V$, we have
\begin{eqnarray}
&&m_{i_{0};x,y}(\mathcal{D}_{i},\mathcal{D}_{e})  \nonumber \\
&\cong &N-\frac{NA_{i_{0}+k}(\hat{q}_{x}^{1}P_{i_{0}+k}^{1}+p_{D})}{C_{i_{0}+k}}
+\overline{h}_{a}^{\prime }(\hat{r}^{1}_{x,y})\sqrt{V}\Phi ^{-1}(2^{-\delta _{1}}) \nonumber \\
&&+\frac{N\hat{q}_{x}^{1}}{C_{i_{0}+k}}\sqrt{%
A_{i_{0}+k}P_{i_{0}+k}^{1}(1-P_{i_{0}+k}^{1})}\Phi ^{-1}(2^{-\delta _{2}}) 
\nonumber \\
&&+\overline{h}_{a}(\hat{r}^{1}_{x,y})\frac{N\hat{q}_{x}^{1}}{C_{i_{0}+k}}\left[
A_{i_{0}+k}P_{i_{0}+k}^{1}\right.   \nonumber \\
&&\left. -\sqrt{A_{i_{0}+k}P_{i_{0}+k}^{1}(1-P_{i_{0}+k}^{1})}\Phi
^{-1}(2^{-\delta _{2}})\right] +\delta _{3}.
\end{eqnarray}
Our task is to choose the parameter $a$ that minimizes $m_{i_{0};x,y}(\mathcal{D}_{i},\mathcal{D}_{e})$; 
the problem is reduced to the minimization of 
$f(a)=\overline{h}_a(\hat{r}^{1}_{x,y})S+ \overline{h}_a^{\prime }(\hat{r}^{1}_{x,y})T$ 
with respect to $a$, 
where 
\begin{align}
f^{\prime }(a)=
\left\{
\begin{array}{cl}
(S(r-a)+T) h^{\prime \prime}(a) & \hat{r}^{1}_{x,y} \ge a, \\
0 & \hat{r}^{1}_{x,y} < a,
\end{array}
\right.
\end{align}
\begin{equation}
S:=\frac{N\hat{q}_{x}^{1}}
{C_{i_{0}+k}} \left[ A_{i_{0}+k}P_{i_{0}+k}^{1}-%
\sqrt{A_{i_{0}+k}P_{i_{0}+k}^{1}(1-P_{i_{0}+k}^{1})}
\Phi ^{-1}(2^{-\delta _{1}})\right],
\end{equation}
and
\begin{equation}
T:=\sqrt{V}\Phi^{-1}(2^{-\delta _{2}}).
\end{equation}
Since $h^{\prime \prime}(a)\le 0$, the minimal $f(a)$ is achieved 
when $a=\hat{r}^{1}_{x,y}+T/S$.
The values of $S$ and $T$ are actually unknown and vary from time to time.
However, if we can expect almost constant values for $S$ and $T$, which can be measured in advance, a favorable choice of $a$ is $a=\hat{r}^{1}_{x,y}+T/S$.

\section{Computation of sacrifice key size (Forward case)}

To compute the size of sacrifice bits in the case of forward error correction, we firstly change the definitions of random variables in Sec.~\ref{General} since $J^{3}$ in Eq.~(\ref{Phase_Error_1}) cannot be expressed in terms of them with their original definitions.
Major alteration concerns the definition of the subscript $j$:
For $j = 0,2,\ldots,2k+1$, $q^{j}$ are changed to represent the detection ratio {\em including} the detector dark counts and $q^{-1}$ is changed to stand for the dark count ratio given that the emitted state is a single photon state ($q^{1}$ is left unchanged).
Namely, for $j=0,2,\ldots,2k+1$, $C_{i}^{j}$ now denote the numbers of pulses detected normally as well as by dark counts given that the emitted state is $\rho _{j}$ and $C_{i}^{-1}$ now stands for the number of dark counts given that the emitted state is $\rho _{1}$.
Alongside the meaning of $r^{j}$ is changed for $j=2,\ldots,k+1$.
According to these alterations, the definition of random variables $E_{i}^{j}$, $F_{i}^{j}$, $G_{i}^{j}$, and $H_{i}$ are also subject to modification.
Almost all of equations in Sec.~\ref{Eve} and Sec.~\ref{Random} are left unchanged except that
(i) $p_{D}$ should read $p_{D} P_{i}^{1}$.
(ii) the range of $q^{j}$ ($j=2,\ldots,2k$) should be changed to $0\le q^{j} \le 1$.
(iii) the range of $x$ should be changed to $[1,\sqrt{2}]$.
and 
(iv) Eqs.~(\ref{q}), (\ref{zeta}), and (\ref{xi}) should read, respectively,
\begin{equation}
\hat{q}^{j}_{x}=\left\{ 
\begin{array}{cl}
p_{0} & j=0 \\ 
\xi ^{j} & j=1,\ldots ,k,k+2,\ldots ,2k \\ 
\frac{1}{\sqrt{2}}\left( x+\xi ^{k+1}\right)  & j=k+1 \\ 
\frac{1}{\sqrt{2}}\left( x-\xi ^{k+1}\right)  & j=2k+1,
\end{array}
\right.
\end{equation}
\begin{eqnarray}
\zeta ^{j} &:=&\frac{1}{\hat{q}_{x}^{j}}\sum_{i=1}^{k}(P_{k\times
k}^{-1})_{j}^{i} \nonumber \\
&&\times \left[ s_{i}p_{i}-\frac{1}{2}%
(P_{i}^{0}p_{0}+P_{i}^{1}P_{D})-P_{i}^{k+1}\hat{q}_{x}^{k+1}y\right],
\end{eqnarray}
and
\begin{equation}
\xi ^{j}:=\sum_{i=1}^{2k}(\overline{P}_{2k\times
2k}^{-1})_{j}^{i}\left( p_{i}-\overline{P}_{i}^{0}p_{0}-\overline{P}%
_{i}^{1}p_{D}-x\overline{P}_{i}^{2k+1}\right).
\end{equation}

The upper bound for the average phase error is now given by
\[
2^{-\left[ m-N+F_{i_{0}+k}^{1}\left( 1-\overline{h}\left( \frac{%
G_{i_{0}+k}^{1}}{F_{i_{0}+k}^{1}}\right) \right) +F_{i_{0}+k}^{0}\right]
_{+}},
\]
Here, we use
\[
N-\frac{NA_{i_{0}+k}}{C_{i_{0}+k}}\left\{ \hat{q}_{x}^{1}P_{i_{0}+k}^{1}%
\left[ 1-\overline{h}_{a}(\hat{r}_{x,y}^{1})\right] +\frac{C_{0}}{A_{0}}%
\right\} 
\]
as the estimation of
\[
N-F_{i_{0}+k}^{1}\left[ 1-\overline{h}\left( \frac{G_{i_{0}+k}^{1}}{%
F_{i_{0}+k}^{1}}\right) \right] -F_{i_{0}+k}^{0}.
\]
The subsequent argument is parallel to that in Sec.~\ref{Reverse}.


\begin{thebibliography}{99}
\bibitem{comment1}
More generally, Eve can choose $q(\mathbf{B})^{j}$ stochastically according to the value of $\mathbf{B}$.
Such an operation can be regarded as the probabilistic mixture of the cases when $q(\mathbf{B})^{j}$ is chosen deterministically.
However, it is sufficient to consider Eve's operation in an extremal point for the security arguments \cite{Hayashi_Tight}.
\bibitem{comment2}
The state $\rho_{-1}$ is regarded as the pulse detected by dark counts.
\bibitem{comment3}
We adopt the linear estimator here.
Otherwise if we adopted the maximal likelihood estimator, 
the computation of variance $v_{i_{0;x,y}}$ would be intractable.
\bibitem{comment4}
If the light source is unstable, for example, the values of $B^{j}_{i}$
	may move somewhat away from the ideal values $A_{i}
	P^{j}_{i}$~\cite{Wang07a,Wang07b}.
In this case, $m_{i_{0};x,y}(\mathcal{D}_{i},\mathcal{D}_{e})$ should be replaced by
\begin{eqnarray*}
&&m_{i_{0};x,y}^{\prime }(B_{i}^{j},\mathcal{D}_{i},\mathcal{D}_{e}) \\
&=&N+\frac{N\left\{ \hat{q}_{x}^{1}B_{i_{0}+k}^{1}\left[ \overline{h}_{a}(%
\hat{r}_{x,y}^{1})-1 \right] -A_{i_{0}+k}p_{D}\right\} }{C_{i_{0}+k}} \\
&&+\sqrt{v_{i_{0;x,y}}(\mathbf{q}_{x,y,(B_{i}^{j})}^{ML},\mathbf{r}%
_{x,y,(B_{i}^{j})}^{ML},B_{i}^{j},\mathbf{A},\pmb{\mu})}\Phi ^{-1}(2^{-\delta
_{1}})+\delta _{3}.
\end{eqnarray*}
Suppose $\mathbf{B}$ belongs to the set $\mathcal{X}$ with probability $1-\epsilon$.
Then, if we define the size of privacy amplification 
$m^{\prime}_{i_{0}}(\mathcal{D}_{i},\mathcal{D}_{e})$
as
\[
m^{\prime}_{i_{0}}(\mathcal{D}_{i},\mathcal{D}_{e})=
\max_{\mathbf{B}\in \mathcal{X},0\le x\le \sqrt{2}(1-p_{D}),0 \le y\le 1}
m^{\prime}_{i_{0};x,y}(B_{i}^{j},\mathcal{D}_{i},\mathcal{D}_{e}),
\]
we have from (\ref{Phase_Error_2}), 
$P^{\mathcal{P}}_{ph,av,\leftarrow}\le\epsilon+2^{-\delta _{1}}+2^{-\delta _{3}} $.
\bibitem{comment5}
Since any photon states can be expanded in terms of coherent states, 
the formalism presented in this paper is still valid for any distributions of photons from the light source by changing the generating probability matices [Eq.~(\ref{Pk})] appropriately.
\end{thebibliography}
\end{document}